\documentclass[USenglish]{article}

\usepackage{authblk}
\usepackage{graphicx}
\usepackage{amsmath}
\usepackage{amsfonts}
\usepackage{braket}
\usepackage{xcolor}
\usepackage[utf8]{inputenc}                           
\usepackage[sort&compress,square,numbers]{natbib}
\usepackage{geometry}
\newcommand{\Ce}[1]{#1}
\newcommand{\T}[0]{\mathcal T}

\author[1]{Andrew Wu \footnote{ These authors contributed equally to this work.} }
\newcommand\CoAuthorMark{\footnotemark[\arabic{footnote}]} 
\author[2]{Javier Cerrillo \protect\CoAuthorMark}
\author[3]{Jianshu Cao\thanks{Corresponding author: \texttt{jianshu@mit.edu}.}} 
\affil[1]{Department of Chemistry, Massachusetts Institute of Technology, 77 Massachusetts Avenue, Cambridge, MA 02139, USA.}
\affil[2]{\'Area de F\'isica Aplicada, Universidad Polit\'ecnica de Cartagena, 30202 Cartagena, Spain.}
\affil[3]{Department of Chemistry, Massachusetts Institute of Technology, 77 Massachusetts Avenue, Cambridge, MA 02139, USA.}
\title{Extracting Kinetic Information from Short-Time Trajectories: Relaxation and Disorder of Lossy Cavity Polaritons}
\date{}

\begin{document}

\maketitle

\begin{abstract}
The emerging field of molecular cavity polaritons has stimulated a surge of experimental and theoretical activities and presents a unique opportunity to develop the many-body simulation methodology. This paper presents a numerical scheme for the extraction of key kinetic information of lossy cavity polaritons based on the transfer tensor method (TTM). Steady state, relaxation timescales and oscillatory phenomena can all be deduced directly from a set of transfer tensors without the need for long-time simulation. Moreover, we generalize TTM to disordered systems by sampling dynamical maps and achieve fast convergence to disordered-averaged dynamics using a small set of realizations. Together, these techniques provide a toolbox for characterizing the interplay of cavity loss, disorder, and cooperativity in polariton relaxation and allow us to predict unusual dependences on the initial excitation state, photon decay rate, strength of disorder, and the type of cavity models. Thus, we have demonstrated significant potential in the use of the TTM towards both the efficient computation of long-time polariton dynamics and the extraction of crucial kinetic information about polariton relaxation from a small set of short-time trajectories.
\end{abstract}

\section{Introduction}

Recent experiments of molecular systems in optical cavities have brought much excitement in chemical sciences, material engineering, and quantum physics,
and have inspired many numerical simulations of  the coherent light-matter interactions in cavity systems.\cite{garcia21,xiong23}
These studies often push the boundary of computational capacities because of the coherent nature of  many-body interactions and the presence of 
various dissipative channels.
\cite{ruggenthaler18,schafer20a,herrera18,ribeiro18,campos23,delpino18,groenhof18,groenhof19,cui22,li22,cao203,sun22,mandal22,cao210}     
In this sense, cavity systems present a unique opportunity to develop and extend the many-body simulation methodology. 
Specifically, early studies have revealed the complex relaxation dynamics of cavity polaritons 
but also opened questions for further research: (i) the interplay of static disorder
and various dissipation channels; (ii) characterization of the relaxation pattern and \Ce{timescale}; (iii) dependence on the initial excitation state;
(iv) comparison of standard polariton models.  In this paper,
we aim to develop novel computational techniques based on the transfer tensor method (TTM) to address these challenging questions numerically.  
 
First introduced in 2014, the TTM is a novel black-box method for extrapolating long-time dynamics from short-time trajectories for any dynamical system
\Ce{with a time-translationally-invariant correlation }\cite{2014}.  \Ce{ As shown in Figure \ref{fig:flowchart}a,}
starting from dynamical maps computed via an input-output analysis, one can construct transfer tensors and obtain the complete information of a reduced quantum system, and can thus propagate its dynamics to arbitrarily long times at an error rate that is independent of the propagation time \cite{doi:10.1080/09500349708231894, Pollock2018tomographically,rosenbach2016efficient}. \Ce{In the field of open quantum dynamics, several methods have emerged recently to compute the longtime dynamics or to predict the steady-state and kinetic
information \cite{doi:10.1063/1.3159671,duan2017zero, hsieh2018, Link23, Cygorek23, Makri21, Makri21a}.
These predictions can be reliable if the calculations are converged, but the propagation to long times can be numerically expensive and can potentially accumulate errors. In comparison, the TTM takes advantage of short time trajectories that are computed with high accuracy and predicts not only the steady state but also relaxation rates. More importantly, the TTM is a black-box technique, which can be applied to any dissipative environment, not limited to Gaussian bosonic baths, and can be combined with any quantum or classical dynamics methods.  In this sense, SMatPI \cite{Makri21, Makri21a} also adopts the concept of transfer tensors and can be regarded as the application of the TTM to QUAPI.}

While the TTM is a powerful tool for reconstructing non-Markovian dynamics from short-time trajectories,  novel technical developments are needed 
in its application to cavity polaritons and other many-body quantum systems.  Specifically, we will explore how to extract kinetic information from
TTM and how to generalize the TTM concept to disordered systems.  The layout of the paper is described as follows: 

In Sec. \ref{sec:model}, we begin with a brief introduction to three cavity Hamiltonians and review the basics of the TTM. 
Specifically, we will introduce the Pauli-Fierz (PF)  Hamiltonian, the Dicke Model, i.e., the coupling of $N$ Two-Level Systems (TLS) to a photonic cavity, and its rotating wave approximation, the Tavis-Cummings (TC) model \cite{https://doi.org/10.1002/qute.201800043}.
In these model Hamiltonians, the reduced density matrix has dimensions $2^N$ by $2^N$, and each transfer tensor has dimensions $4^N$ by $4^N$: the computation thus grows exponentially with the size of the system. To resolve this issue, we adopt the numerical technique to use the full identity matrix of the system as the initial condition to learn the short-time quantum dynamics in a single learning trajectory.

While the transfer tensors contain all the dynamic information, they have been used primarily as propagators. 
In Sec. \ref{sec:rate},  we present our approach to directly extract crucial kinetic information including the steady-state
and relaxation rate from the transfer tensors without propagating the density matrix. Further analysis suggests
    a numerically robust approach to identify oscillatory modes and their decay rates. \Ce{ Beyond rates, we can also evaluate
    high-order moments using transfer tensors and this characterize the deviation from the single exponential decay. }
     Together, these techniques provide a tool box 
    for characterizing the polariton relaxation profile and predict  the dependences on the initial state, system size, and photon decay rate.  

   Though the TTM is conceptualized for a given Hamiltonian, its generalization to disordered systems is of broad interest but has not been explored. 
   In Sec. \ref{sec:disorder}, we explore a novel application of the TTM to disordered systems by incorporating 
   the random distribution of system parameters into the dynamical maps.    In this way, \Ce{the disorder-averaged TTM (i.e., DA-TTM)
   is capable of predicting the averaged  relaxation behavior of the disordered system } 
   without repeated density matrix propagation for an extended simulation time. 
  In fact, the  \Ce{DA-TTM} can even converge faster than directly averaging the density matrices over the realizations, 
  since it can extrapolate averaged results from a relatively small sample size. We apply the approach 
   to predict the average relaxation behavior of disordered cavity polaritons 
   and analyze its dependence on the strength of disorder and on the symmetry of the initial state.
   
In Sec.~\ref{sec:DPF}, we compare the three standard cavity models: the Pauli-Fierz (PF)  Hamiltonian, the Dicke Model, the Tavis-Cummings (TC) model \cite{https://doi.org/10.1002/qute.201800043}. We show that, in the weak coupling limit, all models converge to similar results, whereas their behaviour diverge in the strong coupling limit. We show that lifting the resonance between cavity and atoms can restore the similarity between all three models.

\section{Cavity Models and Transfer Tensor Method\label{sec:model}}
\subsection{Cavity Polariton Hamiltonians}

The Dicke model is widely used in quantum optics for describing a variety of physical phenomena, such as superradiance \cite{Emary_2003}. The standard Dicke model describes the dynamics of $N$ TLSs coupled to a single-mode cavity \cite{PhysRev.93.99} with Hamiltonian
\begin{equation}
    H_D = \hbar\omega_ca^{\dagger}a+\hbar\omega_a\sum_{j=1}^{N}\sigma_j^{z}+\hbar g(a+a^{\dagger})\sum_{j}\sigma_j^x,
    \label{eq:DM}
\end{equation}
where $\omega_c$ is the frequency of the cavity, $\omega_a$ is the frequency of each of the identical TLS, and $g$ is the coupling constant between the cavity and the TLS. The operator $a$ is the annihilation operator of the photonic cavity and the $\sigma_j$ operator is the Pauli operator for the $j$th TLS. Note that while the number of cavity levels is physically infinite, to make calculations tractable, we truncate the cavity basis to $n$ levels, so that the annihilation operator $a$ is simply the rank $n$ operator.

In the weak light-matter coupling regime, we often adopt the Tavis-Cummings (TC) model, which applies the rotating wave approximation (RWA) to the Dicke model and omits the counter-rotating terms in the light-matter interaction Hamiltonian \cite{PhysRev.170.379}. This approximation is valid in particular whenever $\omega_c=\omega_a\gg g$. In this case, the Dicke Hamiltonian reduces to:
\begin{equation}
    H_{TC} = \hbar\omega_ca^{\dagger}a+\hbar\sum_{j=1}^{N}\left(\omega_a\sigma_j^{z}+ ga\sigma_j^++ga^{\dagger}\sigma_j^-\right).
    \label{eq:TC}
\end{equation}

In the strong light-matter coupling regime, we often use a generalization of the Dicke Model, known as the Pauli-Fierz (PF) Hamiltonian\cite{power59}
\begin{equation}
H_{PF}=H_D+\hbar\frac{g^2}{\omega_c}\left(\sum_j\sigma_j^x\right)^2,
\label{eq:PF}
\end{equation}
where the term added to the Dicke Hamiltonian $H_D$ is a dipole self-energy (DSE) that accounts for cavity-mediated interactions between TLS 
of the form $\sigma_j^x\sigma_k^x$.  

It is worth noting that the transfer tensor method below is not affected by the choice of the cavity Hamiltonian, 
and can produce the dynamics of any model with identical computational effort. To simplify the analysis and validation of results below,
we will present most of the numerical calculations using the TC model and will compare the three models in Sec. \ref{sec:DPF}. 

Beyond the coherent dynamics, we will consider the effect of a lossy cavity by means of a master equation
\begin{equation}
    \dot{\rho}_f = -\frac{i}{\hbar}[H, \rho_f]+\frac{\kappa}{2}\left(2a \rho_f a^{\dagger}-a^{\dagger}a\rho_f-\rho_f a^{\dagger}a\right),
\label{Liouvillian}
\end{equation}
where $\rho_f$ is the density matrix of the full system composed by the $N$ TLS and the cavity and $\kappa$ is the decay rate of the cavity. We often seek to estimate the relaxation rates of systems as a function of $\kappa$, fixing $\omega_c=\omega_a=0$. As part of our simulations we also examine a variety of initial states: the singly and multi-excited manifolds of $N$ TLS. For example, the singly excited initial state for 3 TLS may be represented as $\ket{\uparrow \downarrow \downarrow}$, and the fully excited as $\ket{\uparrow \uparrow \uparrow}$ or $\ket{N/2,N/2}$. \Ce{Concerning the initial state of the light, we restrict ourselves throughout to the vacuum state.}

\subsection{Transfer Tensor Method}

\Ce{The simulation of the full cavity photon and matter system $\rho_f$ features an unfavorable scaling that can be partly mitigated if the description is reduced to the matter subsystem alone. In general, the associated reduced density matrix $\rho$ will feature non-Markovian and strong coupling effects that preclude the use of a simple time-local master equation. Further, different models of light-matter interactions define different 
non-Markovian features, which can characterize the light-matter entanglement.  
Here we propose the application of a general-purpose method to resolve
this issue.}

\begin{figure}
    \centering
    \includegraphics[width=0.5\columnwidth]{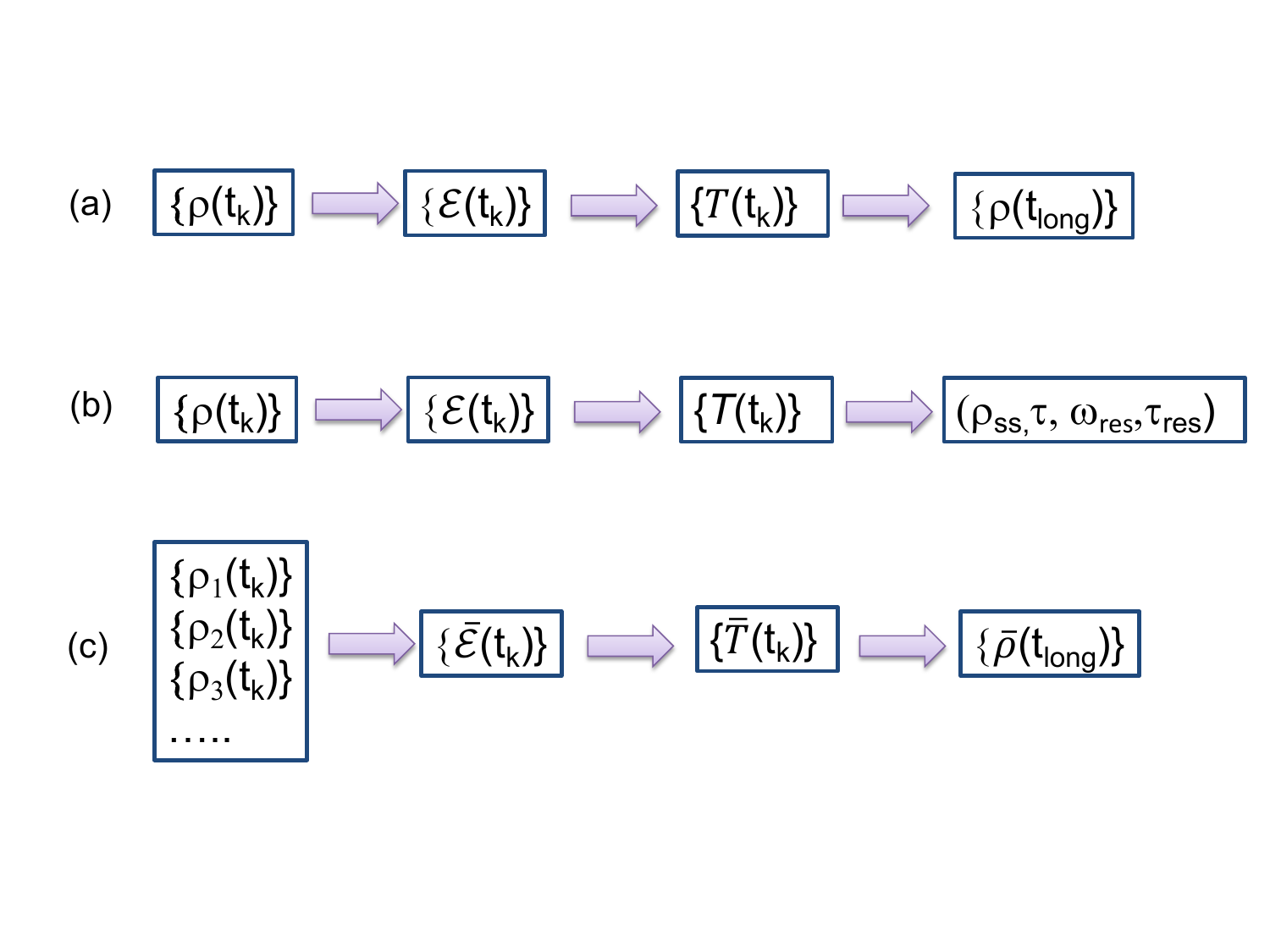}
    \caption{\Ce{Flow charts of TTM and its generalizations: (a) the original TTM, where $\{\rho(t_k)\}$ represents a set of short-time density matrices 
    in the learning period, $\{ \mathcal E(t_k)\}$  and $\{ \mathcal T(t_k)\}$  are the corresponding dynamical maps and transfer tensors, respectively,
    and $\{\rho(t_{long})\}$ represents a set of long-time density matrices predicted by TTM; (b) TTM-based kinetic analysis, where the last step 
   exploits the transfer tensor to extract kinetic information such as the steady-state $\rho_{ss}$, decay lifetime $\tau$, resonance frequency $\omega_{res}$ and its decay time $\tau_{res}$; (c) 
   DA-TTM,  where a disordered ensemble of short-time density matrices is used to generate the disorder-averaged dynamical maps
    and corresponding transfer tensors, which then directly yield the disorder-averaged long-time dynamics or kinetics.}}    
    \label{fig:flowchart}
\end{figure}

First introduced in 2014, the transfer-tensor method (TTM) provides a computational process for extrapolating the complete dynamics of a quantum system from short time trajectories \cite{2014}. \Ce{ As shown in Figure \ref{fig:flowchart}a,}
transfer tensors are extracted from the dynamical maps of the system $\mathcal{E}_k$ associated with a uniform discretization of time $t_k=k\delta t$, with $\delta t$ the time step of the discretization. Dynamical maps are defined by their action on the initial density matrix of the system
\begin{equation}
    \rho(t_k) = \mathcal{E}_k\rho(0),
\end{equation}
that is, the $k$th dynamical map propagates the initial density matrix to the $k$th density matrix according to the dynamics. By definition, $\mathcal{E}_0$ is the identity superoperator acting on system operators.
With the dynamical maps, one can then perform a linear transform to compute the transfer tensors via:
\Ce{
\begin{equation}
    \T_{k} = \mathcal{E}_k-\sum_{m=1}^{k-1}\T_{k-m}\mathcal{E}_m,
    \label{TTM}
\end{equation}}
with \Ce{$\T_1=\mathcal{E}_1$}. Finally, the transfer tensors are used to propagate the dynamics of the system, following the relation:
\Ce{
\begin{equation}
    \rho(t_k)=\sum_{m=0}^{k-1}\T_{k-m}\rho(t_m).
    \label{TTMProp}
\end{equation}}
As pointed out in \cite{2014}, the success of this approach relies on the time-translational invariance of the underlying dynamics. This is guaranteed when (i) the total Hamiltonian is time independent, (ii) the initial total state is a product state, and (iii) the initial environment state is stationary. 
In this paper, one will find that (i)-(iii) all hold for our study of the cavity models. It has been shown that condition (ii) may be relaxed \cite{buser2017initial}.

Further, in the limit $\delta t\rightarrow 0$, the TTM  can be directly connected to the Nakajima-Zwanzig quantum master equation, which allows for generic formulation of open quantum processes \cite{10.1143/PTP.20.948, doi:10.1063/1.1731409}:
\begin{equation}
    \dot{\rho}(t)=-i\mathcal{L}_0\rho(t)+\int_{0}^{t}\mathcal{K}(t-t')\rho(t')dt'.
    \label{eq:NZE}
\end{equation}
In this sense, the transfer tensors are viewed as containing both the Liouvillian, $\mathcal{L}_0$, and the memory kernel, $\mathcal{K}(t)$. That is, the first transfer tensor \Ce{$\T_1$} encodes precisely the Liouvillian via the relation \Ce{$\T_1=(1-i\mathcal{L}_0 \delta t)$}. Similarly, the other transfer tensors \Ce{$\T_i$}  for $i\geq 2$ encode the convolutional memory kernel via the simple relation \Ce{$\T_k = \mathcal K(t_k)\delta t^2$}. In total, this can be summarized with the following relation:
\Ce{\begin{equation}
    \T_{k} = (1-i\mathcal{L}_0\delta t)\delta_{k,1}+\mathcal{K}(t_k) \delta t^2.
    \label{eq:discrete}
\end{equation}}
Although the above expression brings the TTM  into the standard quantum master equation 
formalism of open quantum systems, the TTM is a self-contained conceptual framework and a general computational strategy. 
In fact, the 3 conditions underlying the TTM are sufficient to deduce the Nakajima-Zwanzig equation without using the projection operator formalism. 
 
Theoretically, the TTM has many applications, namely as a dynamic propagator whose accuracy is not determined by propagation time, but rather only by its learning time \cite{rosenbach2016efficient, cao165, duan2017zero}. Analysis has shown TTM is especially promising in scenarios where the propagation time is much longer than the correlation time of the environment \cite{doi:10.1063/1.5009086}.

\subsection{Demonstration of TTM}

In this section, we demonstrate elementary computational results from applying the TTM directly to the TC model, i.e., the RWA of the Dicke Model. For this demonstration, as well as future computations, we take advantage of QuTiP, a python package for simulating quantum systems \cite{JOHANSSON20121760}.

To apply the TTM, one must begin by computing the requisite dynamical maps from short-time dynamics for the $N$ TLS. This may be done by considering the time-local Liouvillian eq.~\eqref{Liouvillian} acting on the density matrix $\rho_f(t)$ of the full system including the cavity
\Ce{\begin{equation}
    \dot{\rho}_f (t)= -i \mathcal L_{f}\rho_f(t).
    \label{LF}
\end{equation}
In practice, we will need to restrict ourselves to a truncation of the cavity to the lowest $n$ Fock levels. For the TC model it is sufficient to consider $n=N+1$. For other models, the particular value of $n$ will depend on the strength of the cavity-matter coupling and in practice needs to be converged for each case.}

The full density matrix $\rho_f$ may be reduced to the density matrix of the $N$ TLS $\rho$ by the action of an appropriate projection superoperator $\mathcal P$
\begin{equation}
    \mathcal P \rho_f(t)= Tr_C\left\{\rho_f(t)\right\}\otimes\rho_C=\rho (t)\otimes\rho_C,
\end{equation}
where $Tr_C$ is the partial trace on the degrees of freedom of the cavity and $\rho_C$ is the initial state of the cavity.

By replacing the initial condition for the solution of eq.~\eqref{LF} with the projected identity superoperator of the full system \Ce{$\mathcal I_f$}
\Ce{\begin{equation}
    \mathcal P \mathcal I_f = n^2 \mathcal E_0\otimes \rho_C,
\end{equation}}
and propagating it to times $t_k$, \Ce{$  \mathcal P \mathcal I_f (t_k)$}, the necessary dynamical maps are simply
\Ce{\begin{equation}
    \mathcal E_k = Tr_C\left\{   \mathcal P \mathcal I_f (t_k)\right\}/n^2,
\end{equation}}
from which the transfer tensors may be derived as usual (eq.\ref{TTM}).

\begin{figure}
    \centering
    \includegraphics[width=0.5\columnwidth]{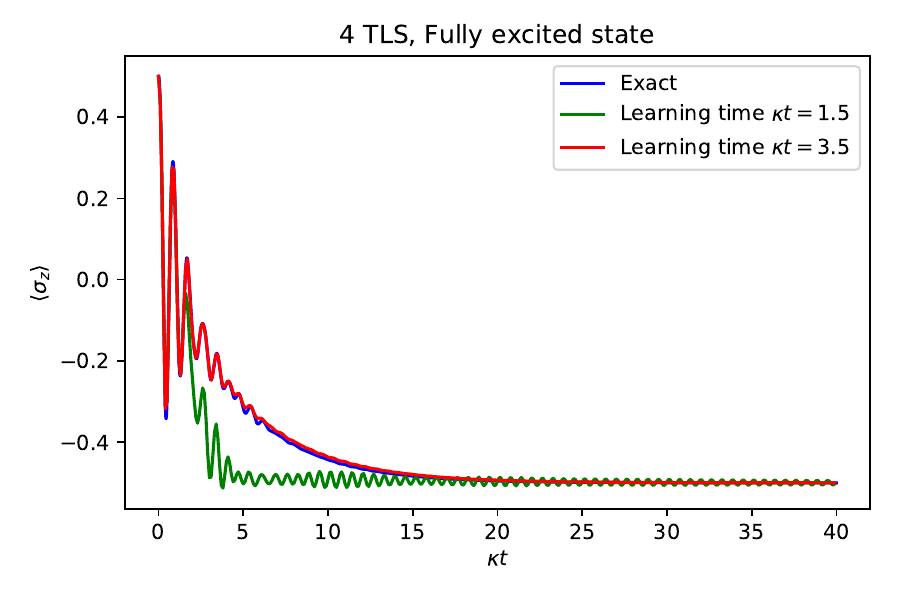}
    \caption{Simulation of the 4-TLS TC Model with a 5-level cavity. The system is parametrized $g=10\kappa/2$, where $\kappa$ is the decay rate of the cavity. Notice that the resonant TC model is simply $H_{TC}=hg(\sigma_{+}a+\sigma_{-}a^{\dagger})$. \Ce{Throughout the manuscript, $\kappa$ or related quantities are taken as the inverse unit of time.} The dynamics are computed for two different learning times, $\kappa t=1.5$ and $\kappa t=3.5$; the latter is found sufficient to accurately capture the reduced dynamics of the system until arbitrary time length.}
    \label{fig:TTM}
\end{figure}

To demonstrate the efficacy of this method, we apply the TTM to a 4-TLS and 5-level cavity TC Model in Figure \ref{fig:TTM}.
Unless otherwise stated, for this and subsequent calculations, we set the cavity and atom frequencies, $\omega_c$ and $\omega_a$ to be on resonance, and set the coupling constant, $g$, to be $10\kappa/\sqrt{N}$, where $N$ is the number of TLS. The only decay channel allowed is via the cavity annihilation operator, with rate $\kappa$.  \Ce{Throughout the manuscript, we take $\kappa$ as a natural unit to define the inverse time and gauge the light-matter coupling. This facilitates the identification of coupling strength regimes.} \Ce{Additionally, we benchmark the TTM results with the exact results 
as computed by direct propagation of master equation \eqref{Liouvillian}.} As shown in the figure, the TTM is very effective at propagating the dynamics of the TC model to times significantly longer than the learning time, although the choice of learning time must be \Ce{reasonable} for the TTM to accurately capture the dynamics. \Ce{In practice, the connection between the transfer tensors and the memory kernel indicates that the learning time must be of the order of the correlation time of the bath.}

\section{Information Extraction via Transfer Tensors \label{sec:rate}}  
\Ce{ As illustrated in Figure \ref{fig:flowchart}b,}
our first main result is to demonstrate how to use the transfer tensors to directly predict key kinetic information without propagation.
 There exist two equivalent paths to demonstrate this:
\begin{enumerate}
\item  direct exploration of the TTM propagation rule eq.~\eqref{TTMProp}
and the unilateral $z$ transform of the transfer tensors as defined by
\Ce{
\begin{equation}
\T\left[z\right]=\sum_{k=1}^{K}z^{-k}\T_k,
\end{equation}
with $K$ the index of the last transfer tensor considered,}
\item or the Laplace transform of the generator of the Nakajima-Zwanzig equation \eqref{eq:NZE}
\Ce{
\begin{equation}
\tilde{\mathcal L}\left(s\right)\equiv -i\mathcal{L}_0+\tilde{\mathcal{K}}\left(s\right),\end{equation}
}
where $\tilde{f}(s) = \int_{0}^{\infty}e^{-st}f(t)dt$ denotes the Laplace transform of the function $f$. 
\end{enumerate}

In the limit of an infinitesimal discretization and by substituting
$z\rightarrow e^{s\delta t}$, the Laplace transform of the generator is recovered from the transfer tensors
\Ce{
\begin{equation}
\tilde{\mathcal L}\left(s\right)=\lim_{\delta t\rightarrow0}\T\left[e^{s\delta t}\right].
\end{equation}
}
which is equivalent to Eq.~\eqref{eq:discrete}.
We show in this section that either perspective allows for the deduction of kinetic information without requiring any propagation of the system.
Beyond this connection, we show in Section \ref{oscill} that further information can be extracted from \Ce{$\T[z]$} as a result of the finite time step $\delta t$. This information is not readily available from \Ce{$\tilde{ \mathcal L}\left(s\right)$} alone.

\subsection{Steady State \label{sec:SS}}
First, we demonstrate the ability to compute the final, infinite-time state (i.e. steady state)  of the non-equilibrium dynamics directly from the transfer tensors. 
We establish the approach first in the continuous Laplace space and then in the discretized transfer tensor form, and finally present an application
to lossy cavities. 

\subsubsection{Continuous Laplace Transform}

To begin, via the Laplace transformation of the quantum master equation,  the density matrix is solved formally as:
\Ce{
\begin{equation}
    \left[s-\tilde{\mathcal L}(s)\right]\tilde{\rho}(s) = \rho(0),
\end{equation}
}
 Then, the steady state can be directly computed via the final value theorem
\begin{equation}
    \rho_{ss} = \lim_{s\rightarrow0} s\tilde{\rho}(s),
\end{equation}
 \Ce{where $ \rho(\infty)=\rho_{ss}$.}
This formalism can be understood 
as the extraction of the overlap between the null subspace of the generator \Ce{ $\tilde{\mathcal L}(0)$} and the initial state $\rho(0)$.

\subsubsection{Discrete Transfer Tensors}

An equivalent representation of the Laplace formalism is provided by the
unilateral z transform of the TTM propagation in eq.~\eqref{TTMProp} as
\Ce{\begin{equation}
\rho\left[z\right]=\T\left[z\right]\rho\left[z\right]+\rho\left(0\right),
\end{equation}}
where we used \Ce{$\T_{0}=0$}. The final value theorem takes the form $\rho_{ss}=\lim_{z\rightarrow1}\left(z-1\right)\rho\left[z\right]$,
so 
\Ce{\begin{equation}
\rho_{ss}=\lim_{z\rightarrow1}\left(z-1\right)\left(1-\T\left[z\right]\right)^{-1}\rho\left(0\right).
\end{equation}}
This result is consistent with the fact that the steady state density matrix must not be affected by propagation with transfer tensors by definition
\Ce{\begin{equation}
\rho_{ss}=\sum_{k}\T_{k}\rho_{ss},
\end{equation}}
which implies \Ce{$(1-\T[1])\rho_{ss}=0$}.  Therefore,  $\rho_{ss}$ is an overlap between the initial state and the null space of \Ce{$1-\sum_k \T_k$}. Alternatively, in the language of complex systems, it can also be interpreted as a fixed point of the transformation represented by the sum of all transfer tensors.

\subsubsection{Application}
\begin{figure}
    \centering
    \includegraphics[width=0.5\columnwidth]{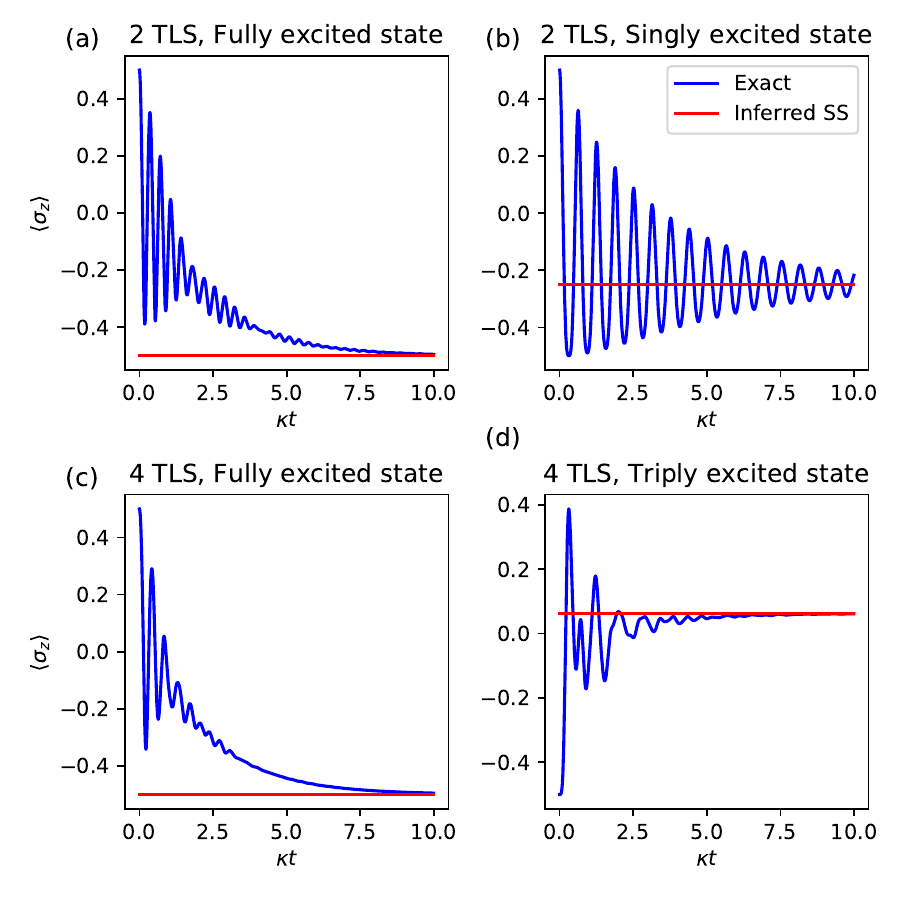}
    \caption{Steady States inferred via the simple matrix calculation from \Ce{Section \ref{sec:SS}}. All diagrams are constructed for the TC Hamiltonian $H_{TC}=\hbar g (\sigma_{+}a+\sigma_{-}a^{\dagger})$ with $g=10\kappa$. Depicted are steady state (red) and initial dynamics (blue) for: a) Fully excited N=2 initial state, b) Singly excited N=2, c) Fully excited N=4, d) Triply excited N=4. Under the partially excited states, since the TC Hamiltonian preserves symmetry, the system is prevented from fully decaying, instead relaxing to a state that remains partially excited.}
    \label{fig:SS}
\end{figure}

As shown in Figure \ref{fig:SS}, we demonstrate the ability of the method to predict the steady states of both the fully excited and partially excited initial states of the TC model. Since this method requires only a handful of matrix arithmetic steps as opposed to fully simulating the system, it can efficiently compute the long-time dynamics of the system with only the short-time trajectories, while remaining agnostic to the ``true'' dynamics of the system.
The steady-state of a lossy cavity depends on the symmetry of the initial state:  The fully-excited initial state relaxes to the ground state 
(i.e., $\langle\sigma_z\rangle=-0.5$), whereas the singly excited initial state or partly excited initial state can remain partly trapped in
 an excited state (i.e., $\langle\sigma_z\rangle \neq -0.5$).

\subsection{Relaxation Rate and Lifetime}

Next, we extract the rate of convergence to the steady state from the transfer tensors. Formally, the relaxation lifetime $\tau$ of a given observable 
$\langle \hat o\rangle$ is defined by the zeroth moment of the time-dependent observable relative to its steady state value, i.e.:
\begin{equation}
   \tau=\frac{\int_0^{\infty}\langle \hat o\rangle(t)-\langle \hat o\rangle(\infty)dt}{\langle \hat o\rangle(0)-\langle \hat o\rangle(\infty)}.
   \label{tau}
\end{equation}
where $\langle \hat o\rangle(\infty) = \langle \hat o\rangle_{ss}$.
Intuitively, this expression can be understood by assuming that $\langle \hat o\rangle(t)$ obeys a simple exponential decay expression of the form 
$\langle \hat o\rangle(t)=\langle \hat o\rangle(\infty)+[\langle \hat o\rangle(0)-\langle \hat o\rangle(\infty)]e^{-t/\tau}$.
 
\subsubsection{Relaxation Matrix}

In this section we generalize the concept of relaxation rate to that of the \emph{relaxation
matrix}
\begin{equation}
\hat{\tau}=\int_{0}^{\infty}\Delta\rho\left(t\right)dt,
\end{equation}
where $\Delta\rho\left(t\right)=\rho\left(t\right)-\rho_{ss}$ is
the deviation of the density matrix from its steady state. For
systems that reach the steady state on a finite time-scale, this matrix
is bounded. By definition, the relaxation matrix is Hermitian and in general not positive,
and depends on the initial condition of the system.

The information contained in the relaxation matrix $\hat\tau$ becomes especially intuitive
when evaluated for the operator of interest $\hat{o}$.  Specifically, we define
\begin{equation}
\left\langle \hat{\tau}\right\rangle _{\hat{o}}=Tr\left\{ \hat{o}\hat{\tau}\right\} =\int_{0}^{\infty}\left[\left\langle \hat{o}\right\rangle \left(t\right)-\left\langle \hat{o}\right\rangle \left(\infty\right)\right]dt,
\end{equation}
which is directly related to the relaxation timescale $\tau$ (eq.~\ref{tau}) via normalization, i.e., 
\begin{equation}
\tau=\frac{\left\langle \hat{\tau}\right\rangle _{\hat{o}}}{\left\langle \hat{o}\right\rangle \left(0\right)-\left\langle \hat{o}\right\rangle \left(\infty\right)}.
\end{equation}

The relaxation matrix $\hat{\tau}$ contains $2^{N}$ real eigenvalues, describing multiple relaxation timescales of the system. 
We may normalize by the initial condition
\begin{equation}
\hat{\tau}'=\left[\Delta \rho(0)\right]^{-}\hat\tau,
\label{eq:relmatnorm}
\end{equation}
where the symbol $-$ implies the pseudo inverse such that just the operator minus its null space is inverted. The normalized relaxation matrix $\hat{\tau}'$
contains a number of non-zero eigenvalues which describe the effective lifetimes of the different
decay modes in the system. In particular, one of the eigenvalues correspond to the depletion timescale from the initial state. 
There exist as many other non-zero eigenvalues as the dimensionality of the null-space that the steady state $\rho_{ss}$ has overlap with. 
These eigenvalues correspond to the timescales needed to reach the steady-state. 
The explicit calculation and spectral analysis of the relaxation matrix will be left for a future study. 

Based on TTM,  we can obtain $\hat\tau$ from direct analysis of the transfer tensors, 
thus avoiding a lengthy and demanding numerical propagation of the density matrix of a system and the subsequent integration.
Below, we will briefly describe the procedure in both the continuous and discrete formalisms.

\subsubsection{Continuous Laplace Transform \label{sec:CLT}}

In analogy to the spectral method presented in
\cite{jung2000,jung1999}, the eigenvalues of \Ce{$\tilde{\mathcal L}(s)$} as a function of $s$ describe all possible
decay (real part) and oscillatory (imaginary part) behaviors of the
system. Further, they contain all the non-Markovian effects
of the bath. Nevertheless, for a system of $N$ TLS,  the analysis of $2^{2N}$ of
these functions of $s$ becomes convoluted. 

Alternatively, the relaxation matrix $\hat\tau$ represents a much more compact object that is directly related
to the Laplace transform of the generator by
\Ce{
\begin{equation}
\hat{\tau}=\lim_{s\rightarrow0}\left[s+\tilde{\mathcal L}\left(s\right)\right]^{-}\left[\rho\left(0\right)-\rho\left(\infty\right)\right],
\end{equation}
}
where again the symbol $-$ implies the pseudo inverse.

\subsubsection{Discrete Transfer Tensors}

Numerically, this calculation can be performed  by means of the z-transform
of the transfer tensors 
\Ce{
\begin{equation}
\hat{\tau}\simeq\lim_{z\rightarrow1}\left(1-\T\left[z\right]\right)^{-}\left[\rho\left(0\right)-\rho\left(\infty\right)\right]\delta t,
\end{equation}
}
and the limit of ${z\rightarrow1}$ simply involves the sum of all transfer tensors $\sum_{k}T_{k}$
\Ce{
\begin{equation}
\hat{\tau}\simeq\left(1-\sum_{k}\T_{k}\right)^{-}\left[\rho\left(0\right)-\rho\left(\infty\right)\right]\delta t.
\end{equation}
}
The transfer tensors are computed based on a time step $\delta t$, and the relaxation matrix  $\hat{\tau}$ converges as $\delta t\rightarrow 0$.

\subsubsection{Application}

\begin{figure}
    \centering
    \includegraphics[width=0.5\columnwidth]{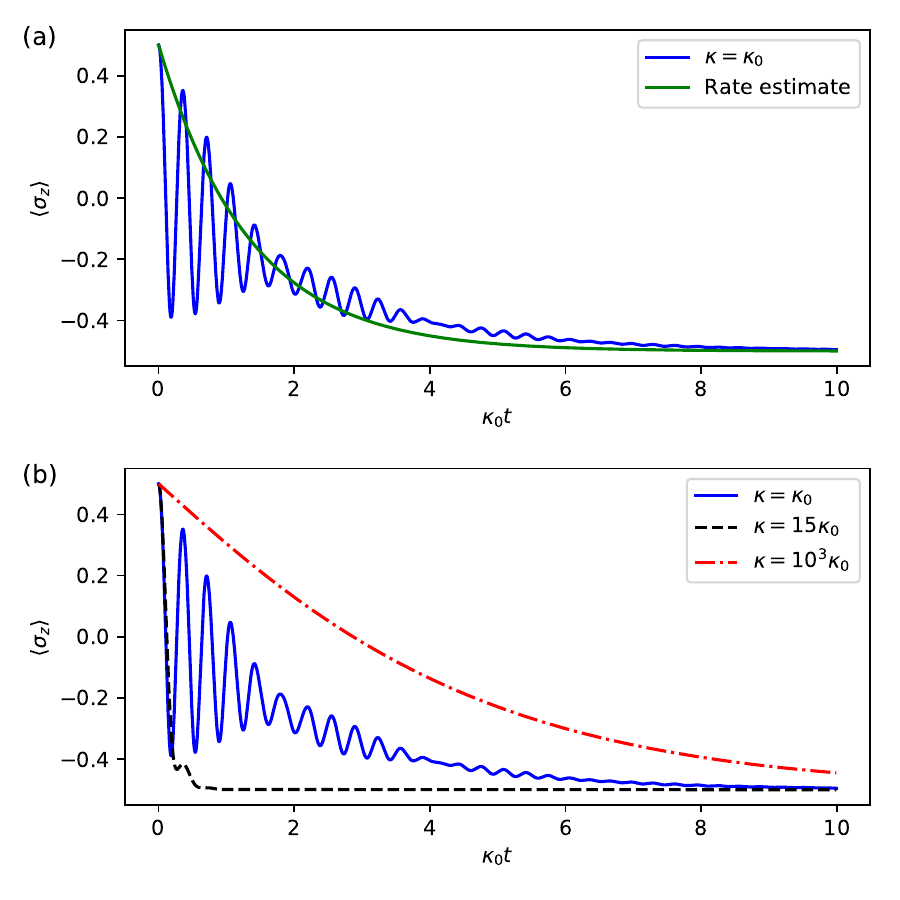}
    \caption{Plot of the relaxation of an individual spin, $\braket{\sigma_z}$, of the 2-TLS, 3-level cavity Tavis-Cummings model with $H=\hbar g(\sigma_{+}a+\sigma_{-}a^{\dagger})$, $g=10\kappa_0/\sqrt 2$\Ce{. (a)} For the underdamped case ($\kappa=\kappa_0$) the estimated relaxation timescale $\tau$ accurately fits the overall dynamics with the function $e^{t/\tau}-1/2$ (green). \Ce{(b) We consider three increasing values of $\kappa$, corresponding to the underdamped ($\kappa=\kappa_0$), critically damped ($\kappa=15\kappa_0$) and overdamped ($\kappa=10^3\kappa_0$) regimes respectively.} }
    \label{fig:test}
\end{figure}

To demonstrate the effectiveness of this method, we consider an initially fully excited $N=2$ TC model with light-matter coupling of $g=10\kappa/\sqrt{2}$ and compute the relaxation timescale $\tau$ of the $\sigma_z$ observable of either TLS. Since each TLS has  initial spin of $1/2$ and steady state spin of $-1/2$, the exponential decay fit takes the form of  $e^{-t/\tau}-1/2$. In Figure \Ce{\ref{fig:test}a}, we plot this exponential decay fit, and find that our method provides a good estimate of the decay rate of the system purely from the transfer tensors.

\begin{figure}
\begin{centering}

\includegraphics[width=0.5\columnwidth]{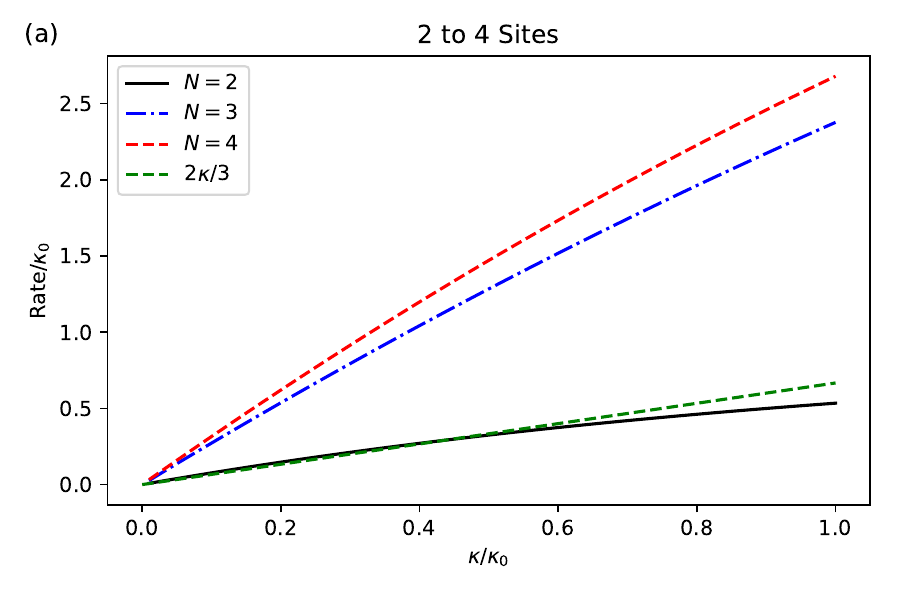}

\includegraphics[width=0.5\columnwidth]{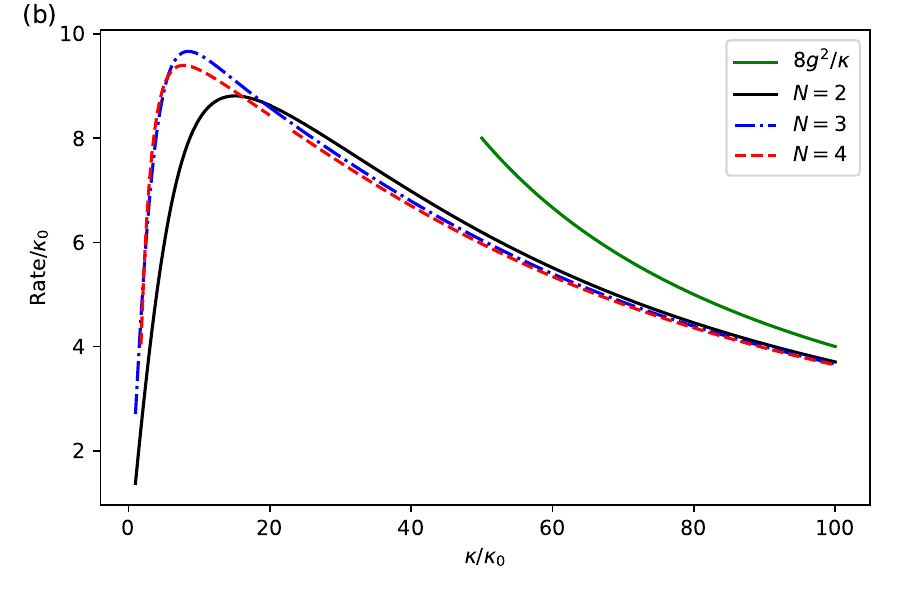}

\caption{\label{fig:Relaxationkappa} Relaxation rate $\tau^{-1}$ \Ce{of the probability of the fully excited state} as a function
of $\kappa$ for $N=2,3$ and $4$ for the fully excited initial state
and $g=10\kappa_0/\sqrt{N}$. (a) For smaller values of $\kappa,$ the
growth is linear. The linear growth for $N=2$ follows the perturbative
scaling $2\kappa/3$. (b) For larger values of $\kappa,$ the
relaxation rate is upper bounded and eventually becomes inversely
proportional to $\kappa$.}
\par\end{centering}
\end{figure}

We now proceed to analyze the relaxation timescale associated with
the dynamics presented in Figure \ref{fig:SS}. 
The operator associated with the relaxation measurement $\hat{o}$ is the projector 
into the fully excited state $\left|N/2,N/2\right\rangle \left\langle N/2,N/2\right|$.
Figure \ref{fig:Relaxationkappa} plots  the inverse
of relaxation timescale (i.e. the relaxation rate) as a function of
the cavity decay rate $\kappa$ of the fully excited initial state \Ce{$\left|N/2,N/2\right\rangle $.}
Three different system sizes of $N$: 2, 3 and 4 are considered.  For sufficiently
small $\kappa$, the rate grows approximately linearly (see Figure \ref{fig:Relaxationkappa}a). The
rate of growth increases with $N$ and, for $N=2$, it coincides with the prediction of perturbation theory of $2\kappa/3$. 
For larger values of $\kappa$, the rate does not follow a trend of
proportionality (see Figure \ref{fig:Relaxationkappa}b). On the
contrary, large cavity decay rates suppress the transfer of excitations
from the atoms to the cavity and reduces the overall effectiveness
of dissipation, to the point where it becomes inversely proportional
to $\kappa$. Thus,  the relaxation rate exhibits a turnover as a function of the cavity decay rate. These correspond to three relaxation regimes illustrated in Figure \Ce{\ref{fig:test}b}.

\subsection{Oscillatory Relaxation \label{oscill}}

In this section we show that, beyond the steady state and relaxation timescales, oscillatory information may be extracted directly from the transfer tensors or the relaxation matrix $\hat \tau$. By analizing their behaviour as a function of $\delta t$ it is possible to make a Fourier transform analysis of the dynamics without actual propagation. We begin by analyzing a case study with a priory knowledge of oscillatory behaviour in Section 3.3.1. We continue in Section 3.3.2 by analytically demonstrating the relationship between the Fourier transform of the dynamics and the analysis of $\hat \tau$ as a function of $\delta t$. Finally, in Section 3.3.3 we show that an enhanced analysis of oscillatory behaviour can be achieved by this method without prior knowledge of the dynamics.

\subsubsection{A Case Study}

\begin{figure}
\centering{}

\includegraphics[width=0.5\columnwidth]{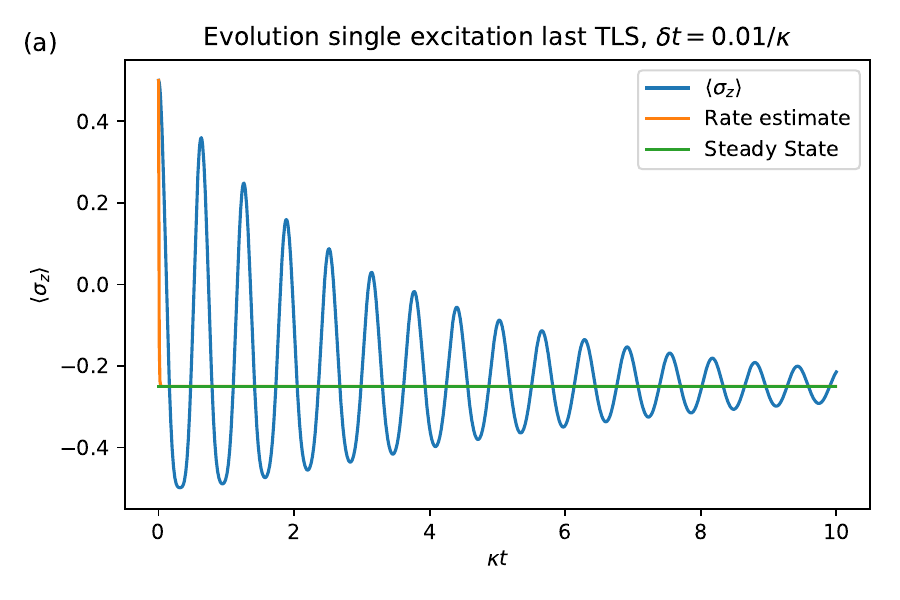}

\includegraphics[width=0.5\columnwidth]{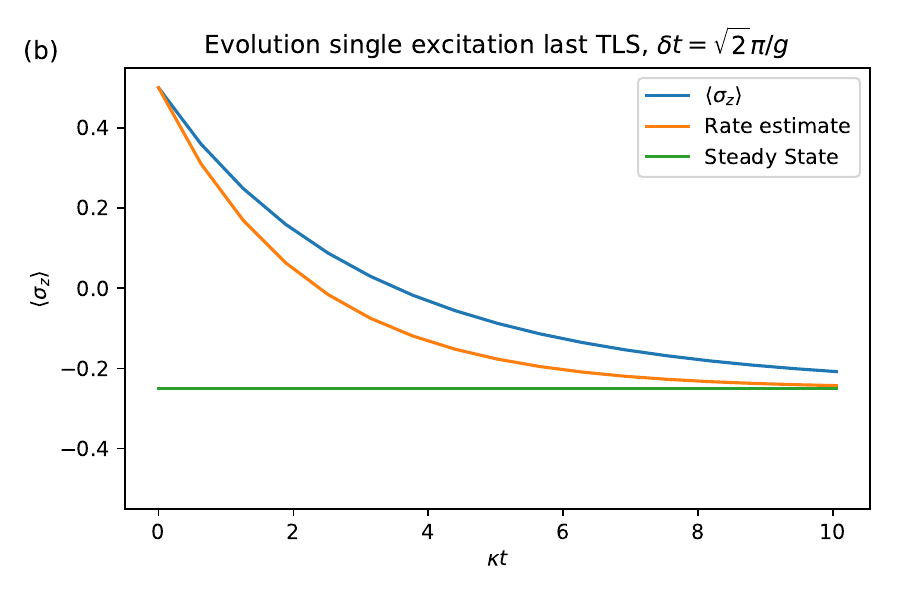}

\caption{\label{fig:Fail}Dynamics of the initially excited TLS $\left\langle \sigma_{z}\right\rangle \left(t\right)$
for $N=2$ (blue). Time is expressed in units of $\kappa^{-1}$. (a) The decay rate estimate $\tau$ in the function
$[3\exp(-t/\tau)+1]/4$ (orange) misses the decay of the envelope.
(b) If the timestep $\delta t$ is adjusted to match the oscillations
observed in the left ($\delta t=\sqrt{2}\pi/g$), just the decaying envelope
is observed by an oversampling effect. The decay rate estimate $\tau$
now approximately matches the decay of the envelope. Parameters
$g=10\kappa /\sqrt{2}$.}
\end{figure}

The estimation of relaxation timescale $\tau$ can further be used to detect oscillatory modes in the system dynamics.
This becomes especially apparent in the relaxation dynamics of a single
excitation initial state. In this case we monitor the operator $\hat{o}=\sigma_{z}$,
the $z$ Pauli operator (i.e., population difference),  of the initially-excited TLS.  As shown
in Figure \ref{fig:SS}b, its dynamics (i.e., population evolution) is oscillatory around the steady state value.
Positive and negative parts of the dynamics cancel each other, so
the overall value of the integral $\tau$ becomes much smaller than
what would be obtained from the decaying envelope and thus does not
constitute an appropriate estimate of the decay rate (see Figure \ref{fig:Fail}a). This can be fixed by adjusting the time-step $\delta t$ of extraction
of the transfer tensors. By tuning $\delta t$  to the period of the oscillations,
a resonance effect takes place by which the decaying envelope
is reproduced and the true relaxation timescale is captured (see Figure
\ref{fig:Fail}b).

\begin{figure}
\centering{}\includegraphics[width=0.5\columnwidth]{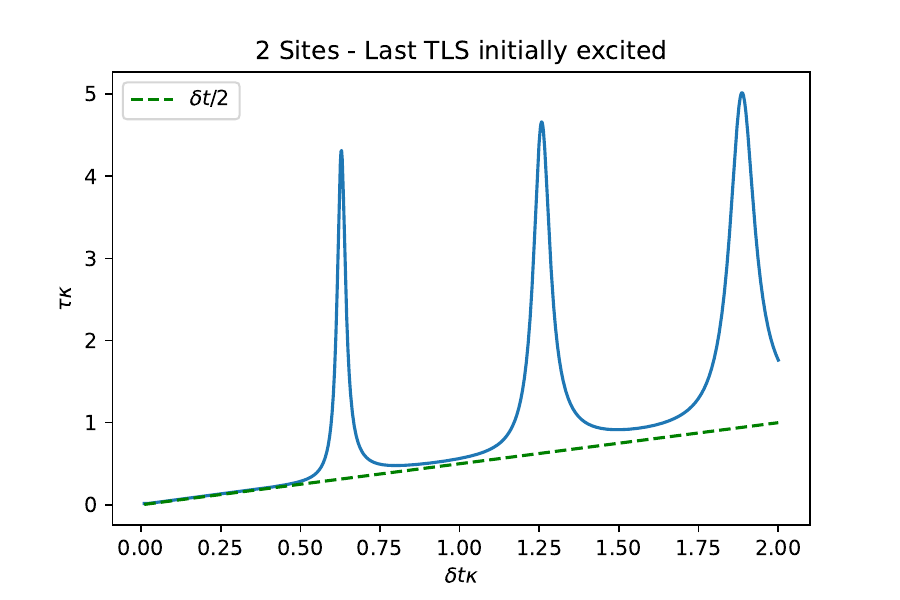}\caption{\label{fig:Comb}Estimate $\tau$ \Ce{for $\sigma_z$} as a function of $\delta t$ for $N=2$
(blue). Clear resonances occur when the timestep $\delta t$ hits the period
of the oscillations observed in Figure \ref{fig:Fail}a, i.e.~$\delta t \kappa=\pi/5$, and its
multiples. As $\delta t$ increases, an error proportional to $\delta t/2$ that
is associated with the discretization of the integral calculation
becomes apparent. Parameters: $g=10\kappa /\sqrt{2}$.}
\end{figure}

In general, an estimate $\tau$ can be computed as a function of
$\delta t$. As $\delta t$ coincides with the period of oscillations of the system or its multiples,
larger estimate $\tau(\delta t)$ is observed. This is
shown in Figure \ref{fig:Comb}, where a resonance is observed once
 $\delta t$ coincides with a multiple of the period of the oscillations
in Figure \ref{fig:Fail}a. An increase of $\delta t$ involves an error
of discretization in the calculation of the time integral, which is proportional
to $\delta t/2$ and can be easily subtracted.

\subsubsection{Resonance Analysis}

The resonance effect as a function of $\delta t$ is tightly connected
to the Fourier transform of 
$\Delta\left\langle \hat{o}\right\rangle \left(t\right)=\left\langle \hat{o}\right\rangle \left(t\right)-\left\langle \hat{o}\right\rangle \left(\infty\right)$.
 Let us explore this connection with an example \Ce{of an arbitrary observable $\langle\hat o\rangle$ featuring an} oscillatory decay described by 
\begin{equation}
\left\langle \hat{o}\right\rangle \left(t\right) = \cos( \omega t) e^{-rt}\left[\left\langle \hat{o}\right\rangle \left(0\right) -\left\langle \hat{o}\right\rangle \left(\infty\right) \right] + \left\langle \hat{o}\right\rangle \left(\infty\right),
\end{equation}
Using transfer tensors with timestep $\delta t$ to estimate $\tau$, we obtain
\begin{equation}
\tau(\delta t) =\sum_k \cos (\omega k\delta t ) e^{-rk\delta t} \delta t
\end{equation}
\begin{equation}
=Re\left\{\frac{\delta t}{e^{\delta t (r+i\omega)}-1}\right\}=\frac{\delta t}{2} \frac{\cos(\omega\delta t) -e^{-r\delta t }}{\cosh(r\delta t)-\cos(\omega\delta t)},
\end{equation}
which is a function with peaks at multiples of the value $\delta t =2\pi/\omega$. 
Thus, the calculation of  $\tau(\delta t)$ offers insight into the Fourier transform of the dynamics.

The effect of the discretization introduced by the TTM
can be formally elucidated by means of a Dirac comb $\Xi_{\delta t}=\sum_{k}\delta\left(t-k\delta t\right)$,
so that
\begin{equation}
\tau(\delta t) =\frac{\int\Xi_{dt}\Delta\left\langle \hat{o}\right\rangle \left(t\right)dt}{\Delta\left\langle \hat{o}\right\rangle \left(0\right)}.
\end{equation}
By the properties of the Dirac comb under Fourier transformation,
we have
\begin{equation}
\int\Xi_{dt}\Delta\left\langle \hat{o}\right\rangle \left(t\right)dt=\frac{1}{\delta t}\sum_{k}\mathcal{F}\left[\Delta\left\langle \hat{o}\right\rangle \right]\left(\frac{2\pi k}{\delta t}\right),
\end{equation}
where $\mathcal{F}\left[f\right]\left(\omega\right)$ is the Fourier
transform of $f\left(t\right)$. Therefore, the estimate $\tau(\delta t)$  provides a sampling of the Fourier transform of
the deviation $\Delta\left\langle \hat{o}\right\rangle $ at the frequencies
$\omega_{k}=2\pi k/\delta t$. Every time one of these frequencies matches
a peak of the Fourier transform, it appears as a peak in $\tau(\delta t)$.
In Figure \ref{fig:Comb}, a single peak in the Fourier transform $\mathcal{F}\left[\delta\left\langle \hat{o}\right\rangle \right]\left(\omega\right)$
at the value $\omega=\sqrt{2}g$ produces a repetition of the peak
at values $\delta t=\sqrt{2}\pi k/g$.

\begin{figure}
\centering{}\includegraphics[width=0.5\columnwidth]{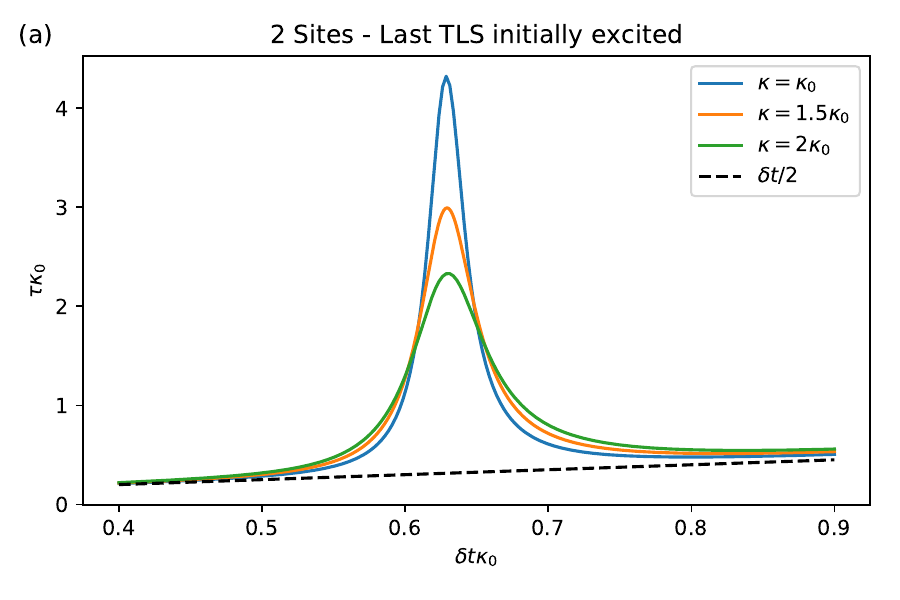}

\includegraphics[width=0.5\columnwidth]{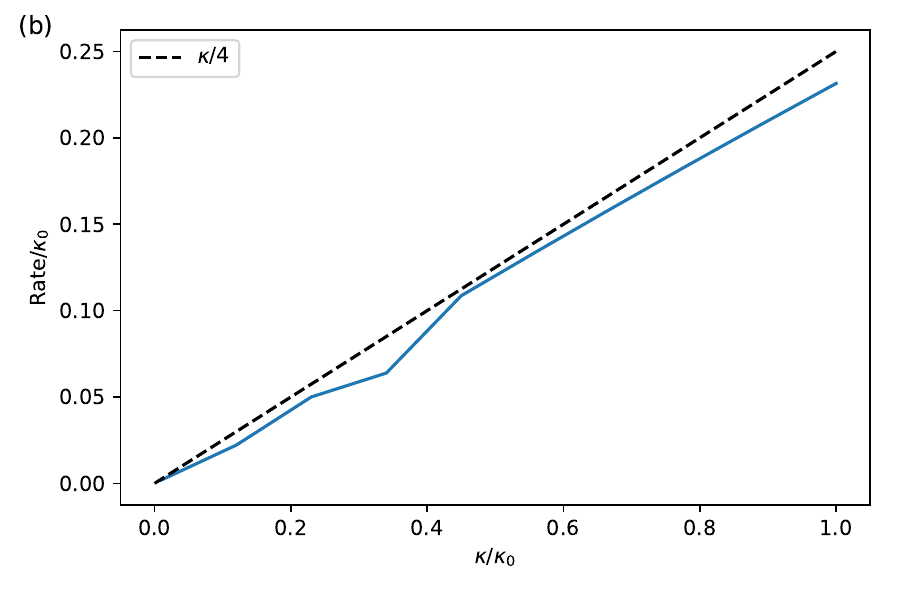}

\caption{\label{fig:kpeak}(a) Estimate $\tau$ \Ce{for $\sigma_z$} as a function of $\delta t$ for
$N=2$ and different values of $\kappa$. As $\kappa$ increases,
the relaxation rate $\tau^{-1}$ at the peak increases too. Parameters:
$g=10\kappa_0/\sqrt{2}$. (b) Rate estimate $\tau^{-1}$ as a function
of $\kappa$ for $N=2$, showing a good agreement with the curve $\kappa/4$.
Parameters: $g=10\kappa_0/\sqrt{2}$.}
\end{figure}

The first peak may be extracted for several values of $\kappa$ as
shown in Figure \ref{fig:kpeak}a. The value of $\tau$ at the
peak (corrected by the error $\delta t/2$) allows us to evaluate the change of  
the relaxation rate as a function of $\kappa$, which
for the case $N=2$ can be analytically proven to follow the linear dependence of $\kappa/4$. 
This is reproduced in Figure \ref{fig:kpeak}b.

\subsubsection{Multiple Resonance}

\begin{figure}
\centering{}
\includegraphics[width=0.5\columnwidth]{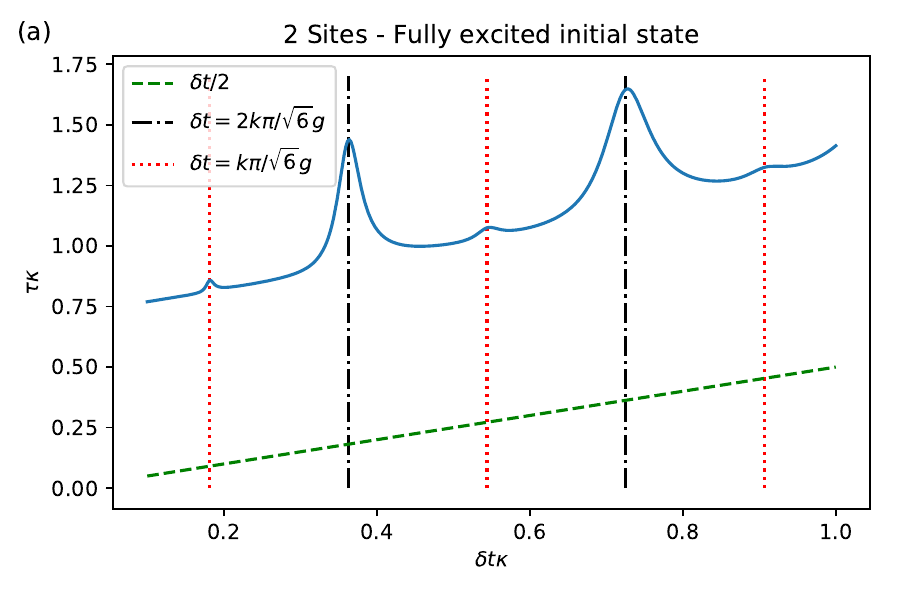}

\includegraphics[width=0.5\columnwidth]{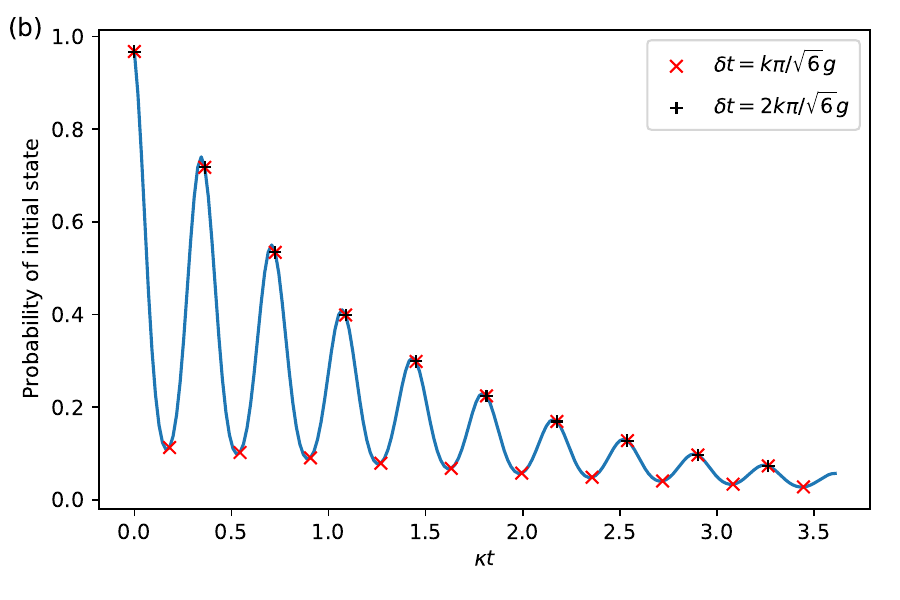}
\caption{\label{fig:Irreg}(a) Estimate $\tau$ \Ce{of the probability of the fully excited state} as a function of $\delta t$ for $N=2$
(blue) and a fully excited initial state. Two types of resonances can be
recognized: one at odd multiples of $\delta t \kappa=\pi/10\sqrt{3}$ (vertical red dotted line) and another one at multiples of $\delta t \kappa=\pi/5\sqrt{3}$ (vertical black dash-dotted line). (b) Probability of the initial state as a function of time. The period of the oscillations matches exactly $\delta t \kappa=\pi/5\sqrt{3}$, as shown by the black plus sign. Half the period is indicated by red crosses. Parameters: $g=10\kappa/\sqrt{2}$.}
\end{figure}

The proposed approach is also
 useful even when the system is initialized in the fully excited state $\left|N/2,N/2\right\rangle $, which does not oscillate around the steady state but may exhibit
an oscillatory decay for small enough $\kappa$. Let us first evaluate the probability of remaining in the initial state for $N=2$, whose dynamics may be analytically solved in the perturbative limit and shown to oscillate at frequency $\sqrt{6}g$. By extracting
the estimate of $\tau$ as a function of $\delta t$, we show in Figure \ref{fig:Irreg}a that two distinct resonances are detected, corresponding precisely to 
$\omega_1=\sqrt{6}g$ and $\omega_2=2\omega_1$. By inspecting the time dependence of the system, Figure \ref{fig:Irreg}a, it becomes clear that $\omega_1$ matches exactly the oscillation frequency of the dynamics and hence maximizes the estimate $\tau$. The double frequency $\omega_2$ matches half the period of oscillation, so both the maxima and minima of the oscillations are sampled. Although the corresponding estimate for $\tau(\delta t)$ cannot be the optimal one, it still provides a higher value than other choices of $\delta t$.

\begin{figure}
\centering{}
\includegraphics[width=0.5\columnwidth]{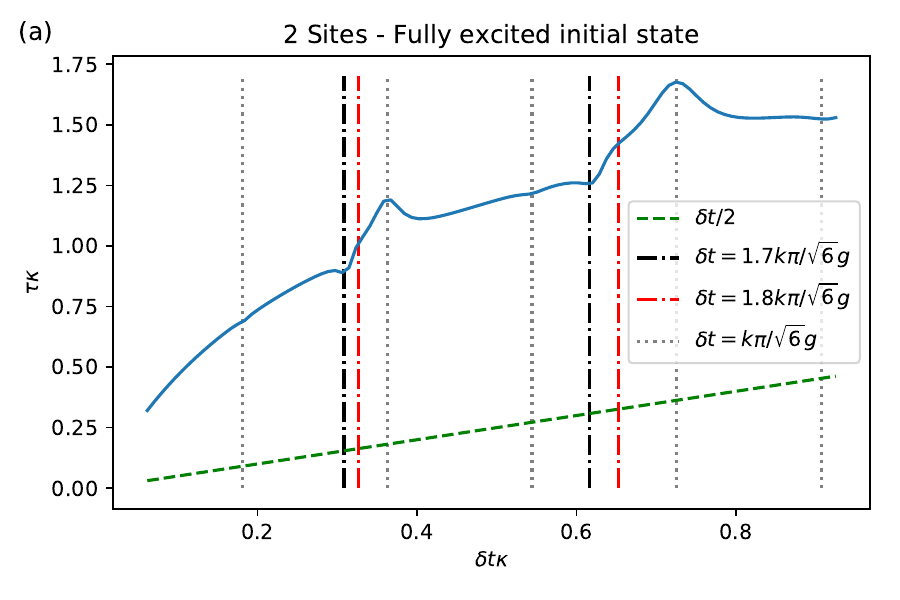}

\includegraphics[width=0.5\columnwidth]{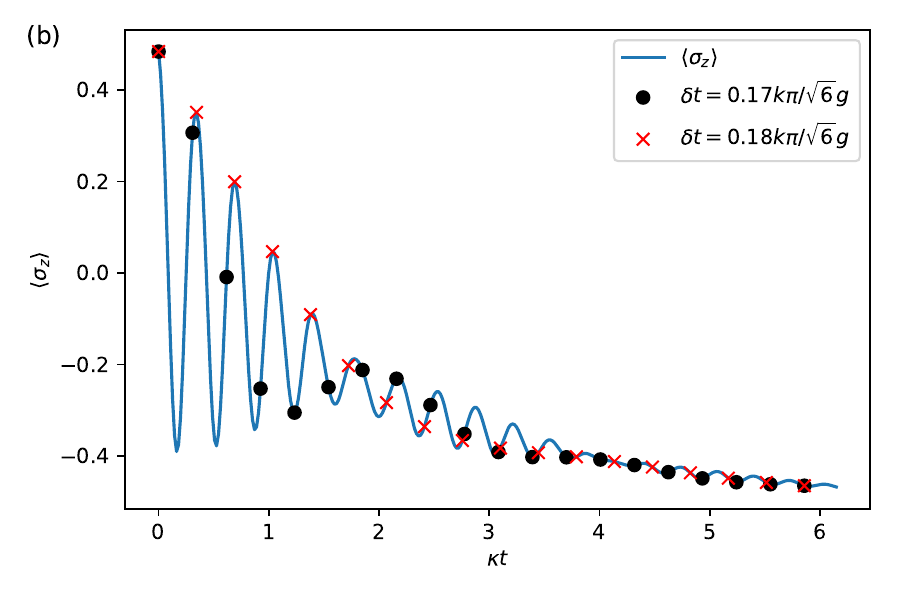}
\caption{\label{fig:Irreg2}(a) Estimate $\tau$ for operator $\sigma_z$ as a function of $\delta t$ for $N=2$ (blue) and a fully excited initial state. Beyond the resonances already present in Figure \ref{fig:Irreg}a (vertical gray dotted lines), two additional resonances can be
observed at $\delta t\kappa=0.17\pi/\sqrt{3}$ (vertical black dash-dotted line) and at $\delta t\kappa=0.18\pi/\sqrt{3}$ (vertical red dash-dotted line). (b) Expected value of operator $\sigma_z$ as a function of time. The two extra resonances roughly correspond to the beating of the dynamics: the first, destructive resonance, corresponds to the half-period of the beating (black dots), whereas the second resonance corresponds to the period of the beating (red crosses). Parameters: $g=10\kappa/\sqrt{2}$.}
\end{figure}

For the same initial state, the expected value of $\sigma_z$ features a more complicated behavior as shown in Figure \Ce{\ref{fig:test}a}. In particular, a beating is apparent between two close frequencies. This pattern shows up in the form of additional resonances in the estimate of the relaxation timescale $\tau$, Figure  \ref{fig:Irreg2}a. Beyond the resonances already identified in Figure  \ref{fig:Irreg}a, 
two more resonances appear that correspond to the beating signal in the dynamics, Figure \ref{fig:Irreg2}b.

In conclusion, the relaxation timescale $\tau$ as a function of the time step $\delta t$ features as a versatile tool to extract relevant oscillatory behavior of the dynamics of open quantum systems.

\Ce{
\subsection{Moments and Poisson Indicator} 
It is known that the Laplace transform of a distribution function is also the moment generating function of time; as such, we can directly calculate all the moments of relaxation via the same formalism. Namely, we have:
\begin{equation}
    M_t(s) = \mathbb{E}[e^{st}] = \tilde{\rho}(-s).
\end{equation}
Thus, we can compute the first moment via:
\begin{eqnarray*}
    M'_t(s) &=& \tilde{\rho}'(-s)\\
    &=&\left[s+\tilde{\mathcal{L}}(-s)\right]^{-2}\left[\mathcal I+\tilde{\mathcal{L}}'(-s)\right]\rho(0).
\end{eqnarray*}
By evaluating this expression in the limit where $s$ approaches 0, one obtains an estimate of $\mathbb{E}[t]$. While mathematically accurate,  this method is numerically unstable, and small variations in either learning time (i.e., the size of the memory kernel) or computation method can lead to varying results. 

However, in order to look at decay dynamics, one must compute the moments of the density matrix 
relative to the steady state density matrix, i.e., $\int_0^{\infty}[t^n\rho(t)-\rho(\infty)]dt$. 
From the derivations in Section \ref{sec:CLT}, we use the relaxation matrix $\hat \tau$ so that the first moment can be computed via:
\begin{equation}
    \hat M'_t(s)=\left[s+\tilde{\mathcal{L}}(-s)\right]^{-2}\left[\mathcal I+\tilde{\mathcal{L}}'(-s)\right]\left[\rho(0)-\rho(\infty)\right],
\end{equation}
where we use the pseudoinverse.
Similarly, we can compute the second moment via:
\begin{eqnarray*}
   \hat M''_t(s) &=&\left\{-2\left[s+\tilde{\mathcal{L}}(-s)\right]^{-3}\left[\mathcal I+\tilde{\mathcal{L}}'(-s)\right]^2+\right.\\
  && \left.\left[s+\tilde{\mathcal{L}}(-s)\right]^{-2}\tilde{\mathcal{L}}''(-s) \right\}\left[\rho(0)-\rho(\infty)\right].
\end{eqnarray*}
Given an operator $\hat o $ of interest, we may project the moment by
\begin{equation}
\langle\hat\tau^2\rangle_{\hat o}=\lim_{s\rightarrow 0} Tr\{\hat o \hat M'_t(s)\},
\end{equation}
and
\begin{equation}
\langle\hat\tau^3\rangle_{\hat o}=\lim_{s\rightarrow 0} Tr\{\hat o \hat M''_t(s)\}.
\end{equation}
With high order moments, we can then characterize the deviation from the exponential decay using the Poisson indicator. The detailed numerical calculation will be left for a future study. 

Since all terms are given entirely by the information in the transfer tensors, we can evaluate this expression for any given short-time simulation from which we can extract transfer tensor information. Thus, given sufficient learning time, this presents a method to tractably compute moments of the dynamics of the system without requiring a full simulation of the system, instead of requiring finite matrix multiplications. With high order moments, we can then characterize the deviation from the exponential decay using the Poisson indicator. The detailed numerical calculation will be left for a future study}

\section{Disorder-averaged \Ce{(DA)} TTM \label{sec:disorder}}

Disordered systems require extensive sampling of initial conditions in numerical simulations. For example,  we  model \emph{disordered} cavity systems by performing simulations on random realizations of the parameters (e.g., $\kappa$, $g$, $\omega_c$, $\omega_a$), drawn from a given probability distribution. 
\Ce{ As illustrated in Figure \ref{fig:flowchart}c,}, we generalize TTM to disordered systems to uncover the disorder-averaged dynamics e.g., 
the effective dynamics by averaging over the random distribution.

Specifically, we consider the TC model with Hamiltonian given by
\begin{equation}
H_{TC} = \hbar\omega_c a^{\dagger} a +\hbar \sum_{j=1}^{N}\left(\omega_{j} \sigma_{j}^{+} \sigma_{j}^{-} +  g a^{\dagger}\sigma_{j}^{-}+ga\sigma_{j}^{+}\right).
\end{equation}
Note that here, as opposed to eq.~(\ref{eq:TC}), each TLS may have its individual frequency $\omega_j$; therefore, it is not possible to remove ${\omega_j}$ uniformly via the rotating wave approximation. To introduce disorder into the system, we fix $g=10\kappa/\sqrt{N}$ (here, $N=2$) and $\omega_c=50\kappa$, but draw $\omega_j/\kappa \in U(40, 50)$.  Below, we use the TTM to extract the disorder-averaged dynamics of the system.

To do so, we explore two techniques: (i) We average the transfer tensors computed for each realization, over all realizations, i.e., 
\Ce{\begin{equation}
\bar{\T} _{k} = \frac{1}{M} \sum_{i=1}^{M} \T_{k}^{i},
\end{equation}}
where $M$ is the number of samples, \Ce{$\bar{\T}_{k}$} is the $k$th transfer tensor used for propagation of the average dynamics, 
and \Ce{$\T_{k}^{i}$} is the $k$th transfer tensor computed from the $i$th realization. (ii) We average the dynamical maps computed via each individual realization, i.e.~
\Ce{\begin{equation}
 \bar{\mathcal{E}}_{k}  = \frac{1}{M} \sum_{i=1}^{M} \mathcal{E}_{k}^{i},
\end{equation} }
where $\bar{\mathcal E}_{k}$ is the average $k$th dynamical map used for computing  the transfer tensor, 
and $\mathcal E_{k}^{i}$ is the $k$th dynamical map computed from the $i$th realization. 
 Below, we compare the result of these two applications of the TTM to the disordered averaged density matrices, which we take to
represent the average behavior of the TLS system.

\begin{figure}
    \centering
    \includegraphics[width=0.5\columnwidth]{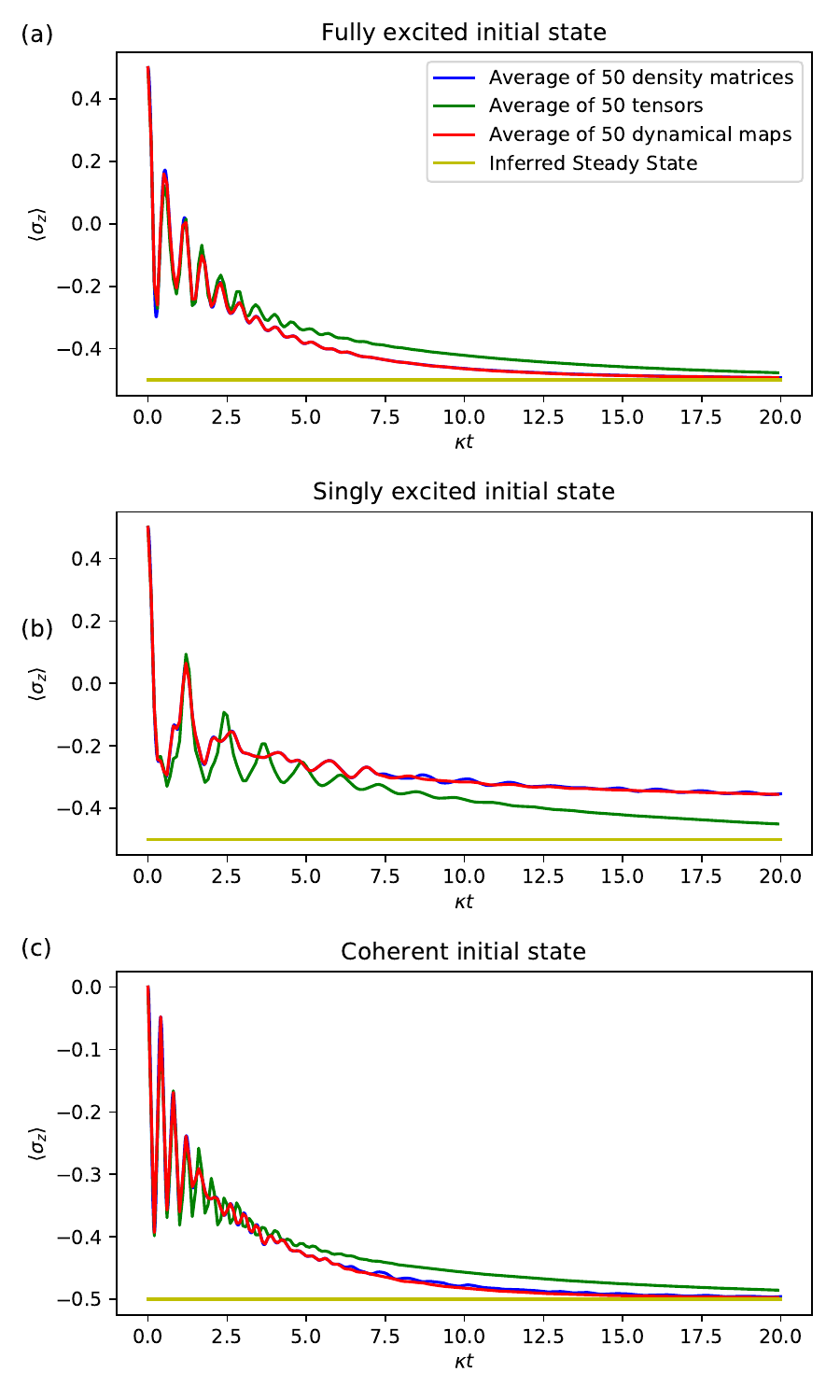}
    \caption{Comparison of simulated spins for various initial states by averaging over 50 realizations. We want to reproduce the behavior resulting from averaging the density matrices of each realization, and two attempts to capture that behavior are given by averaging the dynamical maps or the transfer tensors. All simulations have disorder given by choosing $\omega_j/\kappa \in U(40, 50)$, for the 2-TLS 3-level cavity TC model with $H=\hbar g(\sigma^ + a+\sigma^- a^\dagger)+\hbar \omega_a \sum_{j=1}^{N} \sigma_j^x+\hbar\omega_c a^{\dagger}a$ and $g=10\kappa/\sqrt{2}$, $\omega_c=50\kappa$.(a) Fully excited state. (b) Singly excited state. In this scenario, the true dynamics of the system are markedly different from those deduced via the disorder TTM, and this can be interpreted as the TTM being unable to recover any coherent effects from asymmetric initializations. (c) Coherent superposition initial state, i.e.~all spins are in the coherent superposition $(\ket{\uparrow}+\ket{\downarrow})/\sqrt{2}$. Once again, averaging the density matrices and transfer tensors give good agreement with both each other and the symmetric Hamiltonian dynamics, just as in the fully excited initial state.Tensors are computed with learning time $\kappa t= 4$ and with 10x lower resolution than the density matrices, leading to the figures not matching exactly in (c).}
    \label{fig:disT}
\end{figure}

\begin{figure}
    \centering
    \includegraphics[width=0.5\columnwidth]{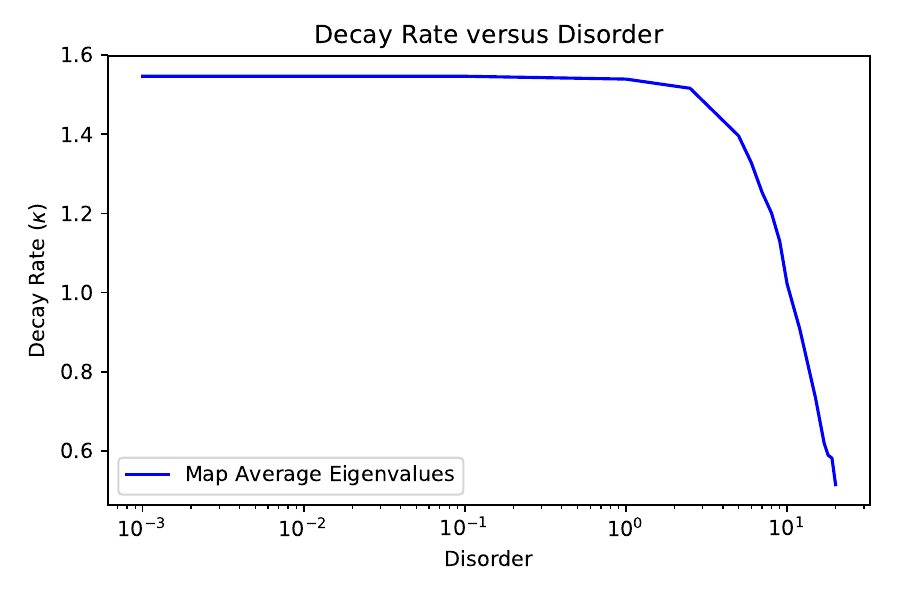}
    \caption{Results from simulating the decay rate as a function of disorder for the 2-TLS Tavis-Cummings Model in the fully excited initial state with $H=hg(\sigma^ + a+\sigma^- a^\dagger)+h\omega_a \sum_{j=1}^{N} \sigma_j^x+h\omega_c a^{\dagger}a$ and disorder given by $\omega_a \in U(45-\delta, 45+\delta)$, $\omega_c=50$ and $g=\frac{10}{\sqrt{2}}$. The decay rate is plotted against log disorder for $\delta \in [0.001, 25]$. As shown, for small magnitudes of disorder there is little impact on the decay rate, but for larger disorders the decay rate quickly decreases.}
    \label{fig:ratesl}
\end{figure}

Figure \ref{fig:disT} summarizes the results of this comparison. As shown in all three plots,  averaging the dynamical maps gives remarkably good agreement with  the averaged density matrices, whereas averaging the transfer tensors exhibits noticeable deviations.  
In addition,  the steady state predicted from the transfer tensors computed via the averaged dynamical maps recover the true, fully-decayed steady state of the system, as expected. Thus, applying the TTM to disordered systems via averaging the dynamical maps computed from each realization of disorder
allows for successful reproduction of disorder-averaged behavior for the system. \Ce{This approach, named DA-TTM', is physically justified, since averaging the dynamical map $\bar{\mathcal E}_k$ is equivalent to  averaging the density matrix of the system at time $t_k$ over disorder, $\bar\rho(t_k)$,
 when applied to initial state $\rho(0)$ common to all realizations. This is not the case for the averaged transfer tensors $\bar{\T}_k$.}

The three plots in Figure \ref{fig:disT} are for different initial conditions: (a) the fully-excited state, (b) a singly excited state, and (c) the coherent superposition state.  In case (a), the initial state and subsequent dynamics is highly symmetric, thus the agreement is nearly perfect. 
Notably, in the asymmetric initial condition of case (b), there are significant fluctuations within the average of the 50 density matrices, 
which can be interpreted as originating from the increased variance in asymmetric initialization. In comparison,
 the \Ce{DA-TTM} yields better convergence, and thus allows us to deduce the averaged behavior of the system using fewer realizations.

Finally, we compute the largest eigenvalue of the normalized relaxation matrix $\hat{\tau}'$ eq.~(\ref{eq:relmatnorm}) to extract the decay rate of our systems as a function of the disorder. For the TC model in the fully-excited initial state with disorder given by $\omega_j\in 45\kappa \pm \delta$, where $\delta$ is the disorder, we vary $\delta$ and compute the decay rates from the transfer tensors extracted via averaging 50 realizations. The result of this numerical experiment is shown in Figure \ref{fig:ratesl}, where the decay rate is plotted as a function of the logarithm of the disorder. For small disorders, the decay rate of the system is slightly reduced, as one may expect--the introduction of small disorder into the cavity frequency is only sufficient to break symmetry. 
However, for higher levels of disorder, the rate of decay decreases rapidly. \Ce{Compared} with 
 our recent analysis of disordered cavities in the thermodynamic limit of the single excitation manifold,\cite{cao204,cao211}  the difference may arise from 
 the fully excited initial state adopted in this calculation.

In short, we have shown that \Ce{the DA-TTM is able to successfully recover the average behavior of a disordered system} through the averaging of dynamical maps, which allow for the computation of transfer tensors that represent the disorder-averaged dynamics. Additionally, applying our technique for extracting the decay rate from the  transfer tensors shows that increasing levels of disorder suppress 
the relaxation of the cavity system. 
Thus, the \Ce{DA-TTM} performs remarkably well for predicting the average behavior of disordered systems, 
especially for the fast convergence to  a single, smooth solution as well as for the deduction of kinetic information
 about the underlying dynamics.

\section{Comparison of Cavity Models \label{sec:DPF}}

We now apply the tools presented in Section \ref{sec:rate} to a comparison of the Dicke model \eqref{eq:DM}, the TC model \eqref{eq:TC},  and  the Pauli-Fierz (PF) model \eqref{eq:PF}.  In this section we first explore the resonant cavity regime, i.e.~$\omega_a=\omega_c=\omega$. While we can take $\omega=0$ in the rotating frame of the TC model, this is not the case in the Dicke or PF models due to the counter-rotating terms and, in addition for the PF model, 
the self energy term.

\begin{figure}
    \centering
    \includegraphics[width=0.5\columnwidth]{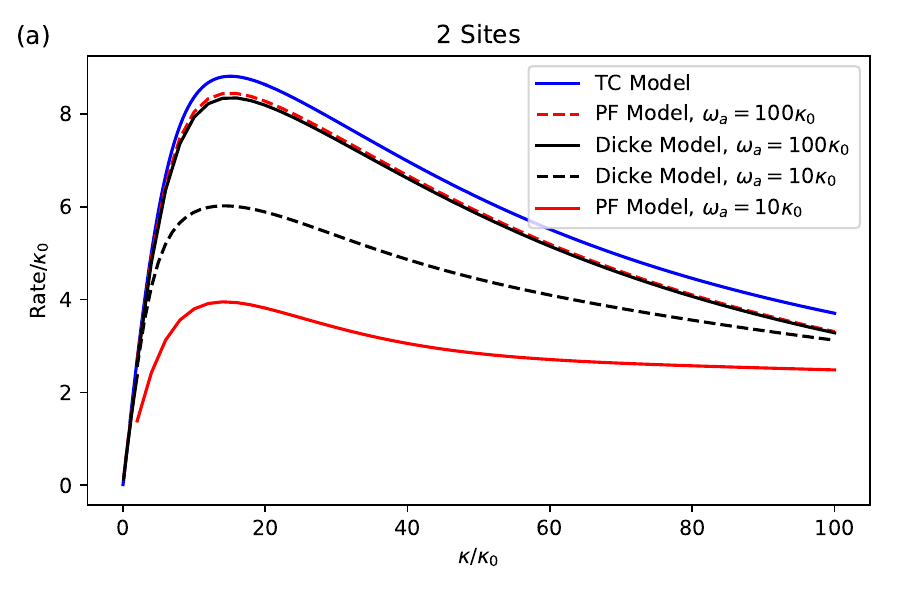}
    
      \includegraphics[width=0.5\columnwidth]{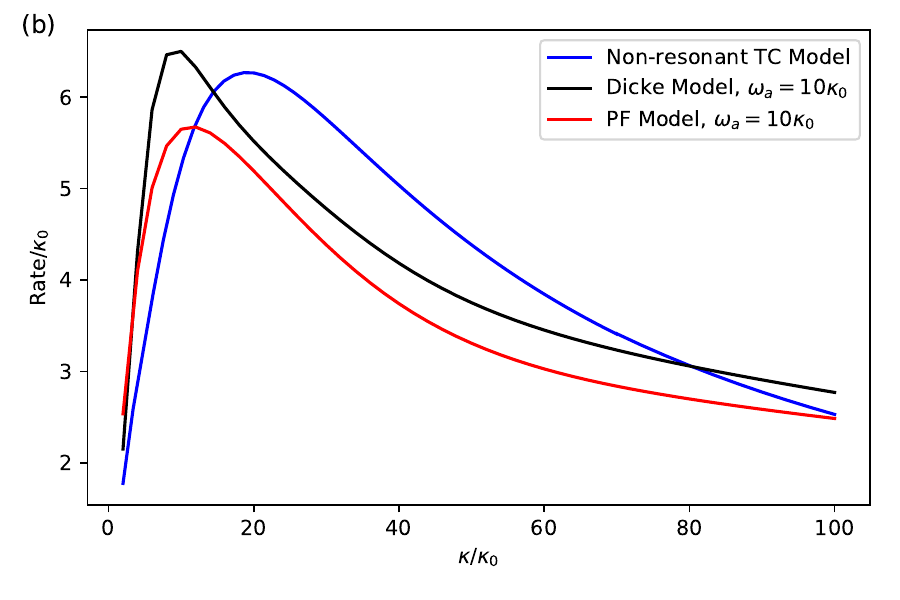}
  
    \caption{Relaxation rate of $\sigma_z$ as a function of cavity decay rate $\kappa$ for three models: TC (equation \ref{eq:TC}) in blue, the Dicke model (equation \ref{eq:DM}) in black and the PF model (equation \ref{eq:PF}) in red. We consider an initially fully excited state and $g=10\kappa_0/\sqrt{2}$. (a) Resonant cavity regime $\omega_a=\omega_c$. For the Dicke and PF model, we consider two values of $\omega_a$ as indicated in the legend. (b) Non-resonant cavity regime with $\omega_a=10\kappa_0$ and $\omega_c=15\kappa_0$.}
    \label{fig:DPFTC}
\end{figure}

In Figure \ref{fig:DPFTC}a we show the relaxation rate $\tau^{-1}$ for operator $\sigma_z$ as a function of the cavity decay rate $\kappa$ for an initially fully excited state. We  consider $N=2$ and two values of $\omega$. The turnover presented in Section \ref{sec:rate} is reproduced in all cases, so both the overdamped and the underdamped limits are explored here.

Let us first discuss the case of $\omega=100\kappa_0$. This corresponds to the regime of a weak cavity-atom coupling $\omega>g$, since $g=10\kappa_0/\sqrt{2}$. In this regime, the counter-rotating terms in both the Dicke and PF models become negligible by virtue of the RWA. Their residual effect, as exposed by the curve corresponding to the Dicke model, is to reduce the overall efficiency of the transfer of excitations from the TLSs and into the cavity. This reduction becomes more critical as the cavity dissipates faster, so the difference between TC and Dicke/PF models becomes apparent only for sufficiently large $\kappa$ (i.e., the overdamped regime). The role of the self energy term in the PF model is also negligible, but it has the ability to partially restore the excitation transfer efficiency. As shown, the relaxation rate of the PF model is slightly larger than that of the Dicke model.

The case of $\omega=10\kappa_0$ corresponds to the strong coupling regime where $\omega\simeq g$. In this limit, neither the counter-rotating terms nor the self-energy term are negligible and they strongly affect the relaxation dynamics in the overdamped limit. In the Dicke limit, a polariton forms between the TLS and the cavity, such that the cavity is displaced from its equilibrium position that is proportional to the spin state of the TLS.\cite{cao192}
This new state relaxes more slowly towards the ground state. In the PF case, the effect of the self energy term is that of shifting and mixing the total spin states of the TLS. Resonance between the cavity and the TLS system is lost, and relaxation rate becomes even further suppressed than in the Dicke model case.

Polariton resonance can be restored in the PF model by increasing the frequency of the cavity mode. This is shown in Figure \ref{fig:DPFTC} (b), where we keep $\omega_a=10\kappa_0$, but increase the cavity frequency to $\omega_c=15\kappa_0$. The PF model increases its relaxation efficiency to a level similar to that of the Dicke model, while the TC model suffers a suppression of its ability to dissipate energy due to the lack of resonance between atoms and cavity. Their qualitative behavior in this limit is very similar, which informs the interpretation that the strong-coupling Dicke model must share traits with an off-resonant TC model and a resonant PF polariton. \Ce{Our observation is consistent with the previous analysis of the Pauli-Fierz model, which displays the change of refractive index in the presence of matter and polarization fluctuations \cite{Ruggenthaler23, Schnappinger23, Sidler23}.}

Recent studies have explored the roles of the dipole self-energy and the rotating wave approximation in these model Hamiltonians.\cite{schafer20a,mandal20,cao217}
Our calculation of the relaxation rate complements these studies, which are mostly based on eigen-solutions and perturbative \Ce{analysis.} 

\section{Conclusion}
In summary, we have developed a novel approach to directly extract kinetic information from the transfer tensors 
without requiring long time propagation and apply the approach to analyze the relaxation process of disordered and lossy cavity polaritons.
 Technically, we have exploited several aspects of the TTM: 
\begin{enumerate}
    \item  The full identity matrix is employed as the initial condition to learn the short-time dynamics in a single learning trajectory.
    \item The null space of the sum of all transfer tensors minus the identity determines the steady state of a given propagation.
    \item The transfer tensors can be transformed into a relaxation matrix, which contains information about decay rates and oscillatory dynamics.
    \item The TTM is also viable for extracting the average behavior of disordered systems via sampling the dynamical maps over realizations.
\end{enumerate}

In particular, we have demonstrated that the information contained in the transfer tensors combined with the initial state of the system is sufficient to compute its corresponding steady state and decay rates. We first applied this technique to a cavity model for a variety of system sizes and initial states, finding that the TTM can accurately predict the long-term equilibrium regardless of the system parameters. Then, we constructed the {\em relaxation matrix} from the transfer tensors, which contains the information about the system's relaxation towards its steady state.  The projection of the relaxation matrix to a particular measurement defines 
the relaxation timescale (i.e., the decay rate) and its high order moments, and the tuning of the relaxation timescale at variable time steps characterizes 
the oscillatory behavior in the relaxation process. 

Equally important is the successful generalization of the TTM to disordered systems. Specifically, the \Ce{DA-TTM} can accurately reproduce disorder-averaged phenomena via averaging the dynamical maps over  realizations of static disorder, thus allowing us to examine  the effects of disorder 
on relaxation kinetics. Further, the \Ce{DA-TTM} can achieve faster convergence than the direct average of the density matrices, 
since it can extrapolate averaged results from a relatively small sample size. 

The application of these novel numerical techniques to polariton relaxation in lossy cavities reveals the rich interplay of disorder, dissipation, 
and cooperativity in light-matter interactions.  Specifically, we have characterized the complex dependence of relaxation kinetics 
 on the initial excitation state, system size, cavity decay rate, strength of disorder, and the type of cavity models. 
 \begin{enumerate}
 \item 
 The steady-state of cavity polaritons depends not only on the equation of motion but also on the symmetry of the initial excitation state:
 The symmetric fully-excited initial state relaxes to the ground state, whereas asymmetric partly-excited initial state will be trapped in 
 an intermediate state without complete relaxation to the ground state. 
 \item
  For the Tavis-Cummings model, the relaxation rate is linearly correlated with the cavity decay rate of the system in the \Ce{weak cavity loss} limit, 
  and the coefficient of the linear dependence depends on the number of TLSs and the initial excitation state.  However, in the \Ce{strong cavity loss} limit,
  the relaxation rate is inversely proportional to the cavity decay rate and is independent of the system size. 
  \item
  The non-monotonic dependence on the photon decay rate defines a turnover which corresponds to the most efficient
  relaxation.  The turnover also signals a transition in the relaxation
  profile from the underdamped oscillations to overdamped decay. 
\item 
 While most studies have been carried out for disordered polaritons in the single excitation manifold, \Ce{our DA-TTM 
  explores the relaxation  of the highly-excited initial state in a disordered cavity and reveals a strong dependence on the initial state.} 
In general, the relaxation rate slightly decreases in the weak disorder regime, as disorder in the cavity frequency is only sufficient to break symmetry,
but drops rapidly in the strong disorder regime.  
\item 
A comparison of standard polariton models, including the Pauli-Fierz (PF)  Hamiltonian, the Dicke model, and the Tavis-Cummings (TC) model, 
reveals a universal turnover as a function of the photon decay rate.
 Though reasonable agreement is achieved in the perturbative and on-resonance regime,
significant differences are observed in the strong light-matter coupling regime which are lifted in the off-resonance case. 
\end{enumerate}

There are, however, multiple routes for significant further progress. For one, 
the TTM has the potential to be combined with any numerical methods for simulating realistic molecular systems in cavities,  
such as ab initial modeling, mixed quantum-classical dynamics, non-adiabatic quantum dynamics. \Ce{\cite{ruggenthaler18, delpino18, groenhof18, groenhof19, cui22, li22,  sun22}}
Additionally, while we have shown the immediate application towards polariton relaxation, our toolbox can be applied generally to
 dissipative quantum processes including quantum transport and quantum thermodynamics,\Ce{with or without couplings to cavity photons.} 

In terms of methodology,  the TTM can be further developed for more efficient analysis of cavity systems.  First, while the decay lifetime  provides an estimator of the relaxation process of the system, it remains to be seen if the time-dependent relaxation matrix can be fully exploited as it contains all the dynamic information. Secondly, by combining symmetry reduction and information extraction from the transfer tensors, we can produce significant computational speed-up. Thirdly, one may consider alternative dimensionality reductions beyond taking advantage of symmetry: e.g., the contraction of the density matrices via an operator, followed by TTM propagation, allowing for prediction of a desirable property with significantly less computation.

\section*{Acknowledgement}
Jianshu Cao acknowledges support from the NSF (Grants No. CHE 1800301 and No. CHE2324300) and from the MIT Sloan Fund.
Javier Cerrillo acknowledges this work is a result of the stay 21734/EE/22 financed by \emph{Fundación S\'eneca-Agencia de Ciencia y Tecnolog\'ia de la Regi\'on de Murcia}\footnote{http://dx.doi.org/10.13039/100007801} through the \emph{Programa Regional de Movilidad, Colaboraci\'on e Intercambio de Conocimiento ``Jim\'enez de la Espada''}. Javier Cerrillo also acknowledges support from Ministerio de Ciencia, Innovación y Universidades (Spain) (‘Beatriz Galindo’ Fellowship BEAGAL18/00078), Grant PID2021-124965NB-C22 funded by MICIU/AEI/10.13039/501100011033 and by “ERDF/EU”, and European Union project C-QuENS (Grant No. 101135359). 

\section*{Author Contributions}
All authors have accepted responsibility for the entire content of this manuscript. Andew Wu and Javier Cerrillo contributed equally to the manuscript. Authors state no conflict of interest.


\begin{thebibliography}{10}

\bibitem{garcia21}
F.~J. Garcia-Vidal, C.~Ciuti, and T.~W. Ebbesen.
\newblock Manipulating matter by strong coupling to vacuum fields.
\newblock {\em Science}, 373:6551, 2021.

\bibitem{xiong23}
Wei Xiong.
\newblock Molecular vibrational polariton dynamics: What can polaritons do?
\newblock {\em Acc. Chem. Res.}, 56(7):776--786, 2023.

\bibitem{ruggenthaler18}
M.~Ruggenthaler, N.~Tancogne-Dejean, J.~Flick, H.~Appel, and A.~Rubio.
\newblock From a quantum-electrodynamical light-matter description to novel
  spectroscopies.
\newblock {\em Nat. Rev. Chem.}, 2:0118, 2018.

\bibitem{schafer20a}
C.~Schafer, M.~Ruggenthaler, V.~Rokaj, and A.~Rubio.
\newblock Relevance of the quadratic diamagnetic and self-polarization terms in
  cavity quantum electrodynamics.
\newblock {\em ACS Photonics}, 7:975, 2020.

\bibitem{herrera18}
F.~Herrera and F.~C. Spano.
\newblock Theory of nanoscale organic cavities: The essential role of
  vibration-photon dressed states.
\newblock {\em ACS Photonics}, 5:65--79, 2018.

\bibitem{ribeiro18}
R.~F. Ribeiro, L.~A. Martnez-Martnez, M.~Du, J.~Campos-Gonzalez-Angulo, and
  J.~Yuen-Zhou.
\newblock Polariton chemistry: Controlling molecular dynamics with optical
  cavities.
\newblock {\em Chem. Sci.}, 9:6325, 2018.

\bibitem{campos23}
Jorge~A. Campos-Angulo, Yong~Rui Poh, Matthew Du, and Joel Yuen-Zhou.
\newblock Swinging between shine and shadow: Theoretical advances on
  thermally-activated vibropolaritonic chemistry (a perspective).
\newblock {\em J. Chem. Phys.}, 158:230901, 2023.

\bibitem{delpino18}
J.~del Pino, F.~A. Y.~N. Schr\"oder, A.~W. Chin, J.~Feist, and F.~J.
  Garcia-Vidal.
\newblock Tensor network simulation of non-Markovian dynamics in organic
  polaritons.
\newblock {\em Phys. Rev. Lett.}, 121:227401, 2018.

\bibitem{groenhof18}
G.~Groenhof and J.~J. Toppari.
\newblock Coherent light harvesting through strong coupling to confined light.
\newblock {\em J. Phys. Chem. Lett.}, 9:4848--4851, 2018.

\bibitem{groenhof19}
G.~Groenhof, C.~Climent, J.~Feist, D.~Morozov, and J.~J. Toppari.
\newblock Tracking polariton relaxation with multiscale molecular dynamics
  simulations.
\newblock {\em J. Phys. Chem. Lett.}, 10:5476--5483, 2019.

\bibitem{cui22}
B.~Cui and A.~Nitzan.
\newblock Collective response in light-matter interactions: The interplay
  between strong coupling and local dynamics.
\newblock {\em J. Chem. Phys.}, 129:173001, 2022.

\bibitem{li22}
Tao~E. Li, Abraham Nitzan, Sharon Hammes-Schiffer, and Joseph~E. Subotnik.
\newblock Quantum simulations of vibrational strong coupling via path
  integrals.
\newblock {\em J. Phys. Chem. Lett.}, 13(17):3890--3825, 2022.

\bibitem{cao203}
Pei-Yun Yang and Jianshu Cao.
\newblock Quantum effects in chemical reactions under polaritonic vibrational
  strong coupling.
\newblock {\em J. Phys. Chem. Lett.}, 12:9531--9538, 2021.

\bibitem{sun22}
Jing Sun and Oriol Vendrell.
\newblock Suppression and enhancement of thermal chemical rates in a cavity.
\newblock {\em JPCL}, 13(20):4441--4446, 2022.


\bibitem{mandal22}
A~Mandal, X~Li, and P~Huo.
\newblock Theory of vibrational polariton chemistry in the collective coupling
  regime.
\newblock {\em J. of Chem. Phys.}, 156:014101, 2022.

\bibitem{cao210}
J.~Cao.
\newblock Generalized resonance energy transfer theory: Applications to
  vibrational energy flow in optical cavities.
\newblock {\em J. Phys. Chem. Lett.}, 13:10943--10951, 2022.

\bibitem{2014}
Javier Cerrillo and Jianshu Cao.
\newblock Non-markovian dynamical maps: Numerical processing of open quantum
  trajectories.
\newblock {\em Physical Review Letters}, 112(11), Mar 2014.

\bibitem{doi:10.1080/09500349708231894}
Isaac~L. Chuang and M.~A. Nielsen.
\newblock Prescription for experimental determination of the dynamics of a
  quantum black box.
\newblock {\em Journal of Modern Optics}, 44(11-12):2455--2467, 1997.

\bibitem{Pollock2018tomographically}
Felix~A. Pollock and Kavan Modi.
\newblock Tomographically reconstructed master equations for any open quantum
  dynamics.
\newblock {\em {Quantum}}, 2:76, July 2018.

\bibitem{rosenbach2016efficient}
Robert Rosenbach, Javier Cerrillo, Susana~F Huelga, Jianshu Cao, and Martin~B
  Plenio.
\newblock Efficient simulation of non-Markovian system-environment interaction.
\newblock {\em New Journal of Physics}, 18(2):023035, 2016.


\bibitem{doi:10.1063/1.3159671}
Keith~H. Hughes, Clara~D. Christ, and Irene Burghardt.
\newblock Effective-mode representation of non-Markovian dynamics: A
  hierarchical approximation of the spectral density. i. Application to single
  surface dynamics.
\newblock {\em The Journal of Chemical Physics}, 131(2):024109, 2009.

\bibitem{duan2017zero}
Chenru Duan, Zhoufei Tang, Jianshu Cao, and Jianlan Wu.
\newblock Zero-temperature localization in a sub-ohmic spin-boson model
  investigated by an extended hierarchy equation of motion.
\newblock {\em Physical Review B}, 95(21):214308, 2017.

\bibitem{hsieh2018}
C.-Y. Hsieh and J. Cao
\newblock 
A unified stochastic formulation of dissipative quantum dynamics. I. Generalized hierarchical equations.
\newblock {\em J. Chem. Phys.} 148(1), 014103/1-14, 2018.

\bibitem{Link23}
Valentin Link, Hong-Hao Tu, Walter T. Strunz
\newblock Open Quantum System Dynamics from Infinite Tensor Network Contraction.
\newblock {\em arXiv preprint}, arXiv:2307.01802, 2023.

\bibitem{Cygorek23}
Moritz Cygorek, Jonathan Keeling, Brendon W. Lovett, Erik M. Gauger
\newblock Sublinear scaling in non-Markovian open quantum systems simulations.
\newblock {\em arXiv preprint}, arXiv:2304.05291, 2023.

\bibitem{Makri21}
Nancy Makri
\newblock Small Matrix Path Integral with Extended Memory.
\newblock {\em J.  Chem. Theory Comput.}, 17(1):1–6, 2021.

\bibitem{Makri21a}
Nancy Makri
\newblock Small Matrix Path Integral for Driven Dissipative Dynamics.
\newblock {\em J.  Phys. Chem. A}, 125(48):10500–10506, 2021.


\bibitem{https://doi.org/10.1002/qute.201800043}
Peter Kirton, Mor~M. Roses, Jonathan Keeling, and Emanuele~G. Dalla~Torre.
\newblock Introduction to the Dicke model: From equilibrium to nonequilibrium,
  and vice versa.
\newblock {\em Advanced Quantum Technologies}, 2(1-2):1800043, 2019.

\bibitem{Emary_2003}
Clive Emary and Tobias Brandes.
\newblock Quantum chaos triggered by precursors of a quantum phase transition:
  The Dicke model.
\newblock {\em Physical Review Letters}, 90(4), jan 2003.

\bibitem{PhysRev.93.99}
R.~H. Dicke.
\newblock Coherence in spontaneous radiation processes.
\newblock {\em Phys. Rev.}, 93:99--110, Jan 1954.

\bibitem{PhysRev.170.379}
Michael Tavis and Frederick~W. Cummings.
\newblock Exact solution for an $n$-molecule---radiation-field hamiltonian.
\newblock {\em Phys. Rev.}, 170:379--384, Jun 1968.

\bibitem{power59}
E.~A. Power and S.~Zienau.
\newblock Coulomb gauge in non-relativistic quantum electro-dynamics and the
  shape of spectral lines.
\newblock {\em Phil. Trans. R. Soc. Lond. A}, 251:427, 1959.


\bibitem{buser2017initial}
Maximilian Buser, Javier Cerrillo, Gernot Schaller, and Jianshu Cao.
\newblock Initial system-environment correlations via the transfer-tensor
  method.
\newblock {\em Physical Review A}, 96(6):062122, 2017.

\bibitem{10.1143/PTP.20.948}
Sadao Nakajima.
\newblock {On Quantum Theory of Transport Phenomena: Steady Diffusion}.
\newblock {\em Progress of Theoretical Physics}, 20(6):948--959, 12 1958.

\bibitem{doi:10.1063/1.1731409}
Robert Zwanzig.
\newblock Ensemble method in the theory of irreversibility.
\newblock {\em The Journal of Chemical Physics}, 33(5):1338--1341, 1960.

\bibitem{cao165}
A.~A. Kananenka, C.-Y. Hsieh, J.~Cao, and E.~Geva.
\newblock Accurate long-time mixed quantum-classical Liouville dynamics via the
  transfer tensor method.
\newblock {\em J. Phys. Chem. Lett.}, 7:4809--4814, 2016.


\bibitem{doi:10.1063/1.5009086}
Andrius Gelzinis, Edvardas Rybakovas, and Leonas Valkunas.
\newblock Applicability of transfer tensor method for open quantum system
  dynamics.
\newblock {\em The Journal of Chemical Physics}, 147(23):234108, 2017.

\bibitem{JOHANSSON20121760}
J.R. Johansson, P.D. Nation, and Franco Nori.
\newblock Qutip: An open-source python framework for the dynamics of open
  quantum systems.
\newblock {\em Computer Physics Communications}, 183(8):1760--1772, 2012.

\bibitem{jung2000}
Jianshu Cao and Younjoon Jung.
\newblock Spectral analysis of electron transfer kinetics. I. Symmetric reactions. 
\newblock {\em The Journal of Chemical Physics}, 112(10), 4716-4722, 2000.

\bibitem{jung1999}
Younjoon Jung, Robert~J Silbey, and Jianshu Cao.
\newblock Electronic coherence in mixed-valence systems: Spectral analysis.
\newblock {\em The Journal of Physical Chemistry A}, 103(47):9460--9468, 1999.

\bibitem{cao204}
G.~Engelhardt and J.~Cao.
\newblock Unusual dynamical properties of disordered polaritons in
  microcavities.
\newblock {\em Phys. Rev. B}, 105:064205/1--19, 2022.

\bibitem{cao211}
Georg Engelhardt and Jianshu Cao.
\newblock Polarition localization and spectroscopic properties of disordered
  quantum emitters in spatially-extended microcavities.
\newblock {\em Phys. Rev. Lett.}, pages 213602/1--7, 2023.

\bibitem{cao192}
M.~Wers\"all, B.~Munkhbat, D.~Baranov, F.~Herrera, J.~Cao, T.~J. Antosiewicz,
  and T.~Shegai.
\newblock Correlative dark-field and photoluminescence spectroscopy of
  individual plasmon-molecule hybrid nanostructures in a strong coupling
  regime.
\newblock {\em ACS Photonics}, 6(10):2570--2576, 2019.

\bibitem{Schnappinger23}
T.~Schnappinger, D.~Sidler, M.~Ruggenthaler, A.~Rubio, and M.~Kowalewski.
\newblock Cavity Born-Oppenheimer Hartree-Fock ansatz: Light-matter properties of strongly coupled molecular ensembles.
\newblock {\em J. Phys. Chem. Lett.}, 14(36):8024-8033, 2023.

\bibitem{Sidler23}
D.~Sidler, T.~Schnappinger, A.~Obzhirov, M.~Ruggenthaler, M.~Kowalewski, and A.~Rubio.
\newblock Unraveling a cavity induced molecular polarization mechanism from collective vibrational strong coupling.
\newblock {\em arXiv preprint}, arXiv:2306.06004, 2023.

\bibitem{Ruggenthaler23}
M.~Ruggenthaler, D.~Sidler, and A.~Rubio.
\newblock Understanding polaritonic chemistry from ab initio quantum electrodynamics.
\newblock {\em Chem. Rev.}, 123(19):11191-11229, 2023.


\bibitem{mandal20}
A.~Mandal, T.~D. Krauss, and P.~Huo.
\newblock Polariton mediated electron transfer via cavity quantum
  electrodynamics.
\newblock {\em J. Phys. Chem. B}, 124:6321, 2020.


\bibitem{cao217}
J.~Cao and E.~Pollak.
\newblock Cavity-induced quantum interference and collective interactions in
  Van der Waals systems.
\newblock {\em arXiv:2310.12881}, 2023.

\end{thebibliography}
\end{document}